\documentclass[sigconf,authorversion]{acmart}

\usepackage[T1]{fontenc}
\usepackage[utf8]{inputenc}

\usepackage{microtype}
\usepackage{booktabs}

\usepackage[english]{babel}
\usepackage{csquotes}

\usepackage{math}
\usepackage{color}
\usepackage{graphicx}
\usepackage{multirow}
\usepackage{subcaption}
\usepackage{xspace}
\usepackage[binary-units=true]{siunitx}
\usepackage[linesnumbered,ruled,vlined]{algorithm2e}
\SetAlFnt{\small\sffamily}

\acmPrice{15.00}

\setcopyright{acmlicensed}

\setcitestyle{square}

\newcommand{\tugaffil}{
	\affiliation{%
		\institution{Graz University of Technology}
		\streetaddress{Inffeldgasse 16}
		\city{Graz}
		\postcode{8010}
		\country{Austria}
	}
}

\newcommand{\figref}[1]{Figure~\ref{#1}}

\newcommand{\eg}{\emph{e.g.}\xspace}
\newcommand{\ie}{\emph{i.e.}\xspace}

\newcommand{\term}[1]{\emph{#1}}
\newcommand{\id}[1]{\texttt{#1}}

\begin{document}

\title{On-the-fly Vertex Reuse for Massively-Parallel Software Geometry Processing}

\author{Michael Kenzel}
\email{michael.kenzel@icg.tugraz.at}
\tugaffil
\author{Bernhard Kerbl}
\email{bernhard.kerbl@icg.tugraz.at}
\tugaffil
\author{Wolfgang Tatzgern}
\email{wolfgang.tatzgern@student.tugraz.at}
\tugaffil
\author{Elena Ivanchenko}
\email{elena.ivanchenko@icg.tugraz.at}
\tugaffil
\author{Dieter Schmalstieg}
\email{dieter.schmalstieg@icg.tugraz.at}
\tugaffil
\author{Markus Steinberger}
\email{markus.steinberger@icg.tugraz.at}
\tugaffil

\renewcommand\shortauthors{Kenzel, M. et al.}

\def\gitorigin{[link removed for review]}

\begin{abstract}
	
	Compute-mode rendering is becoming more and more attractive for non-standard rendering applications, due to the high flexibility of compute-mode execution.
	These newly designed pipelines often include streaming vertex and geometry processing stages.
	In typical triangle meshes, the same transformed vertex is on average required six times during rendering.
	To avoid redundant computation, a post-transform cache is traditionally suggested to enable reuse of vertex processing results.
	However, traditional caching neither scales well as the hardware becomes more parallel, nor can be efficiently implemented in a software design.
	We investigate alternative strategies to reusing vertex shading results on-the-fly for massively parallel software geometry processing.
	Forming static and dynamic batching on the data input stream, we analyze the effectiveness of identifying potential local reuse based on sorting, hashing, and efficient intra-thread-group communication. 
	Altogether, we present four vertex reuse strategies, tailored to modern parallel architectures. 
	Our simulations showcase that our batch-based strategies significantly outperform  parallel caches in terms of reuse.
	On actual GPU hardware, our evaluation shows that our strategies not only lead to good reuse of processing results, but also boost performance by  $2-3\times$ compared to na\"ively ignoring reuse in a variety of practical applications.
\end{abstract}



\begin{teaserfigure}
	\centering
	\begin{subfigure}{0.3\linewidth}
		\centering
		\includegraphics[height=3cm]{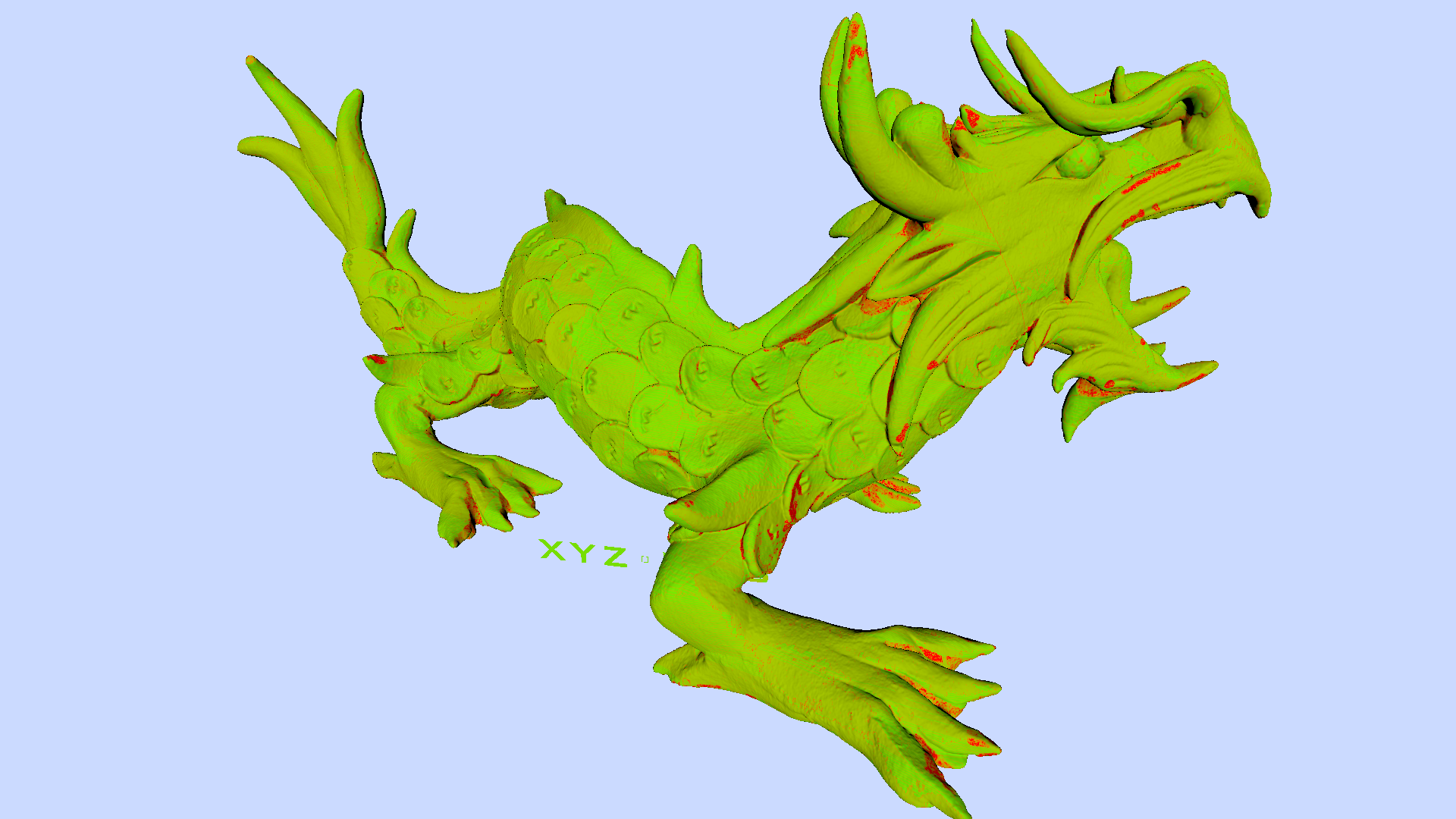}
		\caption{Shading rate for XYZRGB Dragon}
	\end{subfigure}
	\hfill
	\begin{subfigure}{0.3\linewidth}
		\centering
		\includegraphics[height=3cm]{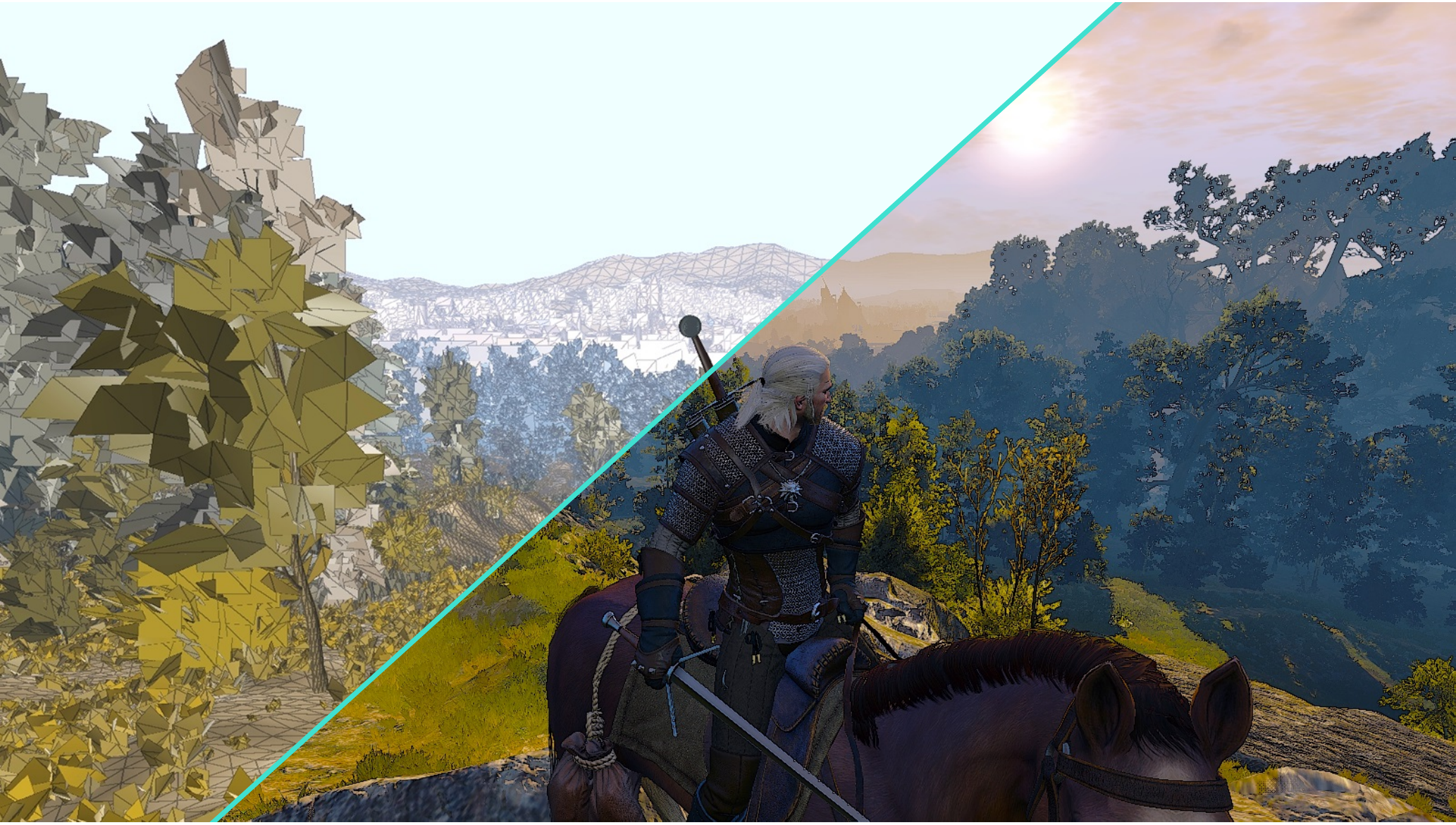}
		\caption{Rendering a scene from \emph{The Witcher 3}}
		\label{fig:game}
	\end{subfigure}\hfill
	\begin{subfigure}{0.39\linewidth}
		\centering
		\includegraphics[height=3cm]{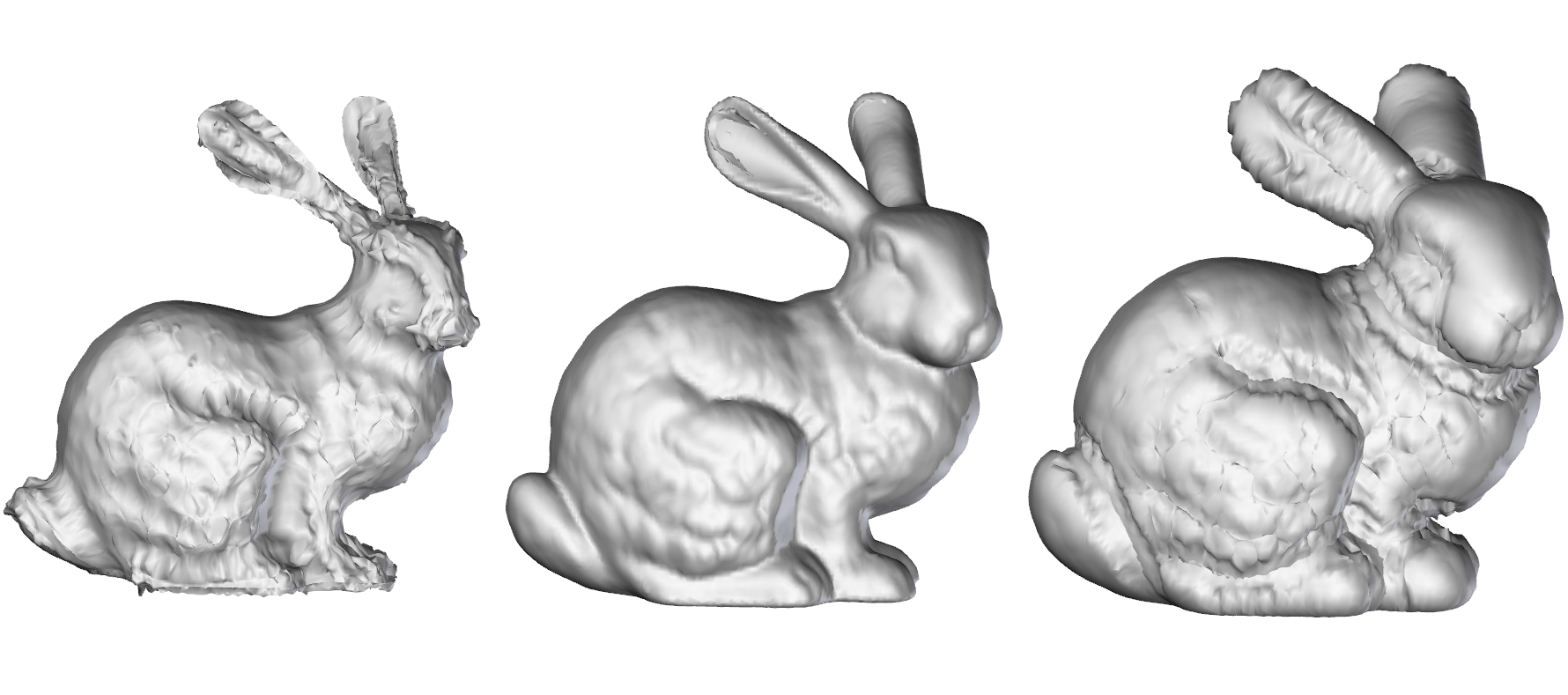}
		\caption{Computation of inner and outer mesh envelopes}
		\label{fig:envelopes}
	\end{subfigure}\hfill
	\caption{
		Reducing the number of shader invocation during rendering is essential to guarantee high performance.
		Traditionally, redundant vertex shading can be bypassed using a post-transform cache, but its poor scalability makes the vertex cache a poor choice in massively parallel environments.
		(a) The batch-based approaches we explore in this work show good reuse characteristics on modern GPUs (green vertices are shaded only once, dark red six times).
		(b, c) We evaluate static and dynamic batching in a variety of applications, \eg, rasterization of captured game scenes and computation of mesh simplification envelopes.
		\emph{The Witcher 3: Wild Hunt screenshot courtesy of CD PROJEKT S.A.; used with permission.}
	}
	\label{fig:testscenes}
\end{teaserfigure}

\maketitle

\section{Introduction}

Although hardware-supported, real-time rendering of 3D scenes is highly efficient, the standard rendering pipeline implemented in hardware lacks flexibility in certain aspects.
With modern graphics processing units (GPU) inexorably rising in compute power, implementing (parts of) custom pipelines in compute-mode, \ie, in software, becomes an interesting alternative.
Although certain features---like rasterization---will likely always be multiple orders of magnitude faster in hardware, others may efficiently be realized also in software.
Primitive transformations, \ie, vertex shading, is one of those applications.
While implementing vertex shading stages in software for execution on GPU compute units becomes more and more common, \emph{vertex reuse}, \ie, reusing the result of the vertex shader when it is referenced more than once, is usually ignored. 
This is in part due to vertex reuse being realized in hardware in the conventional pipeline---a feature that is not exposed for custom use in software.

However, vertex reuse should not be neglected, as it offers several benefits for high-performance rendering. 
In addition to a significant reduction of required memory for storing input geometry data, effective vertex reuse can greatly reduce the number of shader invocations. 
A vertex in a mesh is, on average, referenced up to six times.
The traditional solution to enable vertex reuse is the employment of a \emph{post-transform cache}.
The post-transform cache stores shaded vertex information, which can then be retrieved instead of computing the same information multiple times~\cite{Sheaffer:2004:FSF:1058129.1058142,Wang:2011:PGS:2019608.2019612}.
The significance of this assumption is underlined by the wide body of research aiming at improving the ordering of vertices in meshes to yield better cache behavior.
%
Unfortunately, there is little publicly available information on the implementation specifics used in current GPUs.
The lack of mention of the post-transform cache in more recent articles~\cite{purcell2010fast, Kubisch:2015:LOT} also raises the question, to which degree the widely accepted preconceptions about vertex reuse still hold.

The adequacy of a central vertex cache in contemporary graphics pipelines and its use in a software pipeline should be questioned, and justifiably so: with the increasing degree of parallelism usually present in modern GPUs, the costs of a post-transform cache can be expected to rise drastically. 
Alternative design choices tailored towards massively parallel devices may circumvent this bottleneck while achieving similar or even better reuse characteristics.
%
In this light, we see large potential benefits by revisiting the problem of efficient vertex reuse with an additional focus on software rendering pipelines.
In search of methods capable of scaling with the massively parallel architecture of current and future GPUs, 
we make the following contributions:
\begin{enumerate}
	\item We investigate batch-based vertex uniquization as an alternative to post-transform caching for achieving reuse.
	\item Next to a na\"ive processing scheme, we discuss four batch-based approaches to identify unique vertices on massively parallel devices.
	\item We evaluate all approaches with respect to their theoretical and practical vertex reuse effectiveness in a variety of computer graphics applications.
\end{enumerate}

\section{Related work}

It has been realized early on that there is significant potential for optimization by minimizing redundancy in an input stream describing mesh geometry.
The pioneering work by \citet{Deering:1995:GC:218380.218391}, \citet{Evans:1996:OTS:244979.245626}, and \citet{Chow:1997:OGC:266989.267103} considered the problem from a data compression point of view.
However, due to this angle of approach, these methods required input geometry to always first be encoded according to some compression scheme which would then be decompressed during processing.

Hoppe et al.~\cite{Hoppe:1999:OML:311535.311565} were the first to explore the use of a $k$-FIFO post-transform \term{vertex cache} to reduce redundant vertex processing on-the-fly during rendering of triangle meshes. They furhtermore presented a set of algorithms that automatically optimize the rendering sequence for a given mesh to maximize utilization of their proposed cache architecture. The downside of their optimization approach is that it requires exact knowledge of the properties of the underlying hardware which are subject to change.
However, their work inspired a long line of followup research improving upon their results.~\cite{LinYu:2006, Sander:2007:FTR:1276377.1276489, Chhugani:2007:GEO:1230100.1230102}
Arguably one of the most impactful works is the architecture-agnostic approach by Forsyth~\cite{forsyth2006linear}.

A more current area of research where we encounter the problem of massively parallel vertex processing is software rendering on the modern GPU. Noteworthy examples of GPU software rendering pipelines include Freepipe~\cite{Liu:2010:FPP}, CUDARaster~\cite{Laine:2011:HSR}, and Piko~\cite{Patney:2015:PFA:2809654.2766973}. They all use the compute mode of the GPU (typically on top of the CUDA~\cite{NVIDIA:CUDA} ecosystem) to implement rasterization and fragment shading, but lack mechanisms for vertex reuse. Freepipe simply executes the vertex shader every time an index is fetched. CUDARaster and Piko run the vertex shader in a preprocessing step on the entire vertex buffer and store the results in global memory. This strategy is wasteful if a substantial portion of the vertices is never referenced or used during rendering, which is a rather common case in practice, \eg, with methods such as level-of-detail or occlusion culling. The need for buffering the entire intermediate output of the geometry stage also leads to excessive memory requirements.

In order to avoid ambiguity in the following sections, we will employ the nomenclature for parallel execution and hardware concepts according to CUDA~\cite{NVIDIA:CUDA}. 
Hence, wave fronts of single-instruction-multiple-data (SIMD) width will be referred to as \emph{warp}. 
Warp divergence indicates the case where threads follow redundant execution paths, since warps advance in lockstep. 
Logical groups of warps that run on the same multiprocessor share a portion of fast local \emph{shared memory} and can easily synchronize will be addressed as \emph{blocks}. 

\section{Vertex reuse strategies}

\begin{figure}
	\centering
	\includegraphics[width=\linewidth]{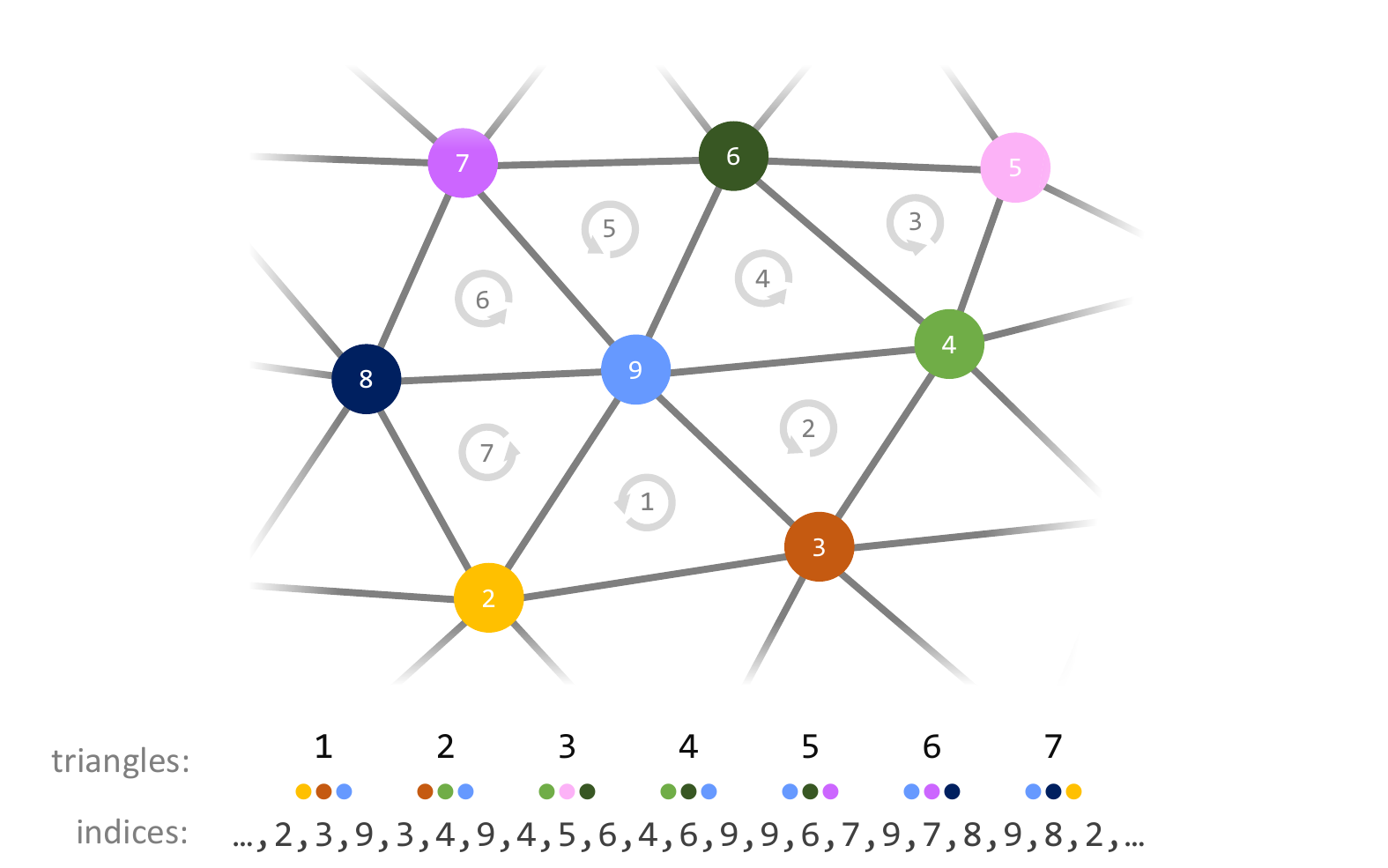}
	\caption{
		Section of a mesh and its representation as an indexed triangle list. On average, each vertex is referenced six times in a typical mesh.
	}
	\label{fig:reuse}
\end{figure}

A major goal of this work is a characterization of vertex reuse in a software-based, massively parallel context, and heightening the understanding of its influence on graphics workload. 
To this aim, we formulate the following assumptions:
We only consider indexed triangles as primitives, for which the index buffer can be used to identify recurring vertices~(\figref{fig:reuse}). 
The routine (or \emph{shader}) for processing a vertex is invoked based on an index buffer, where groups of threads are assigned to consecutive primitives in the index buffer. 
In the ideal case, the vertex shader should be executed only once for each vertex that is referenced by the index buffer. 
To ensure high performance, shading must happen in parallel, without any need for expensive synchronization or communication across GPU multiprocessors. 
In order to support maximum performance in streaming pipelines, preprocessing of the vertex or index buffer should be kept at a minimum, or avoided altogether.

\subsection{The post-transform cache revisited}

Given the above considerations for geometry processing in a streaming pipeline, a global persistent post-transform cache appears to be the intuitive choice to reduce the number of vertex shader invocations. 
However, such a cache is difficult to implement efficiently if it is required to work across  the multiprocessors on the GPU.
On the other hand, even in a single multiprocessor, the high level of data concurrency may defeat the purpose of caching.
The reuse of vertex information can occur almost instantly, if neighboring triangles are referenced in quick succession in the index buffer (\eg, triangle strip layout).
Consequently, an advancing wave front of threads may process the same vertex multiple times in parallel, and new cache entries become available too late to be of use.
In a software-only implementation, caching additionally suffers from high latency when using conventional memory rather than dedicated cache hardware.
Furthermore, it is prone to cause detrimental thread divergence, since cache hits and misses lead to different control paths in the execution of warps~\cite{Clarberg2013}. 

\subsection{Batch-based vertex reuse}
To avoid the issues raised by the use of a central cache, we propose the concept of \emph{batch-based} vertex processing, which naturally lends itself to execution on massively parallel architectures. 
A batch is defined by us as a bounded region in the index buffer, that is assigned to a single warp or block for processing.
The thread block is responsible for executing the shader once for each referenced vertex within its batch and assembles the output triangles. 
Each block must analyze its batch, assign vertices uniquely to threads for shader invocation and finally distribute shading results for assembling the output triangles. 
This implies that duplicate indices in the batch need to be identified before executing the vertex shader.

\begin{figure}[h!]
	\centering
	\begin{subfigure}{\linewidth}
		\includegraphics[width=\linewidth]{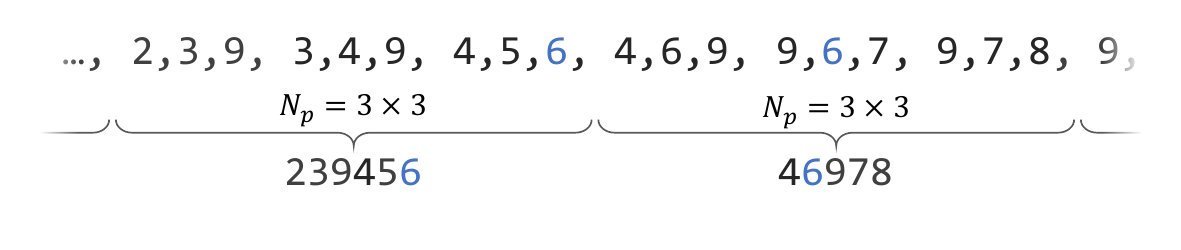}
		\vspace{-23pt}
		\caption{static batching}
		\vspace{10pt}
		\label{fig:static_batching}
	\end{subfigure}
	\begin{subfigure}{\linewidth}
		\includegraphics[width=\linewidth]{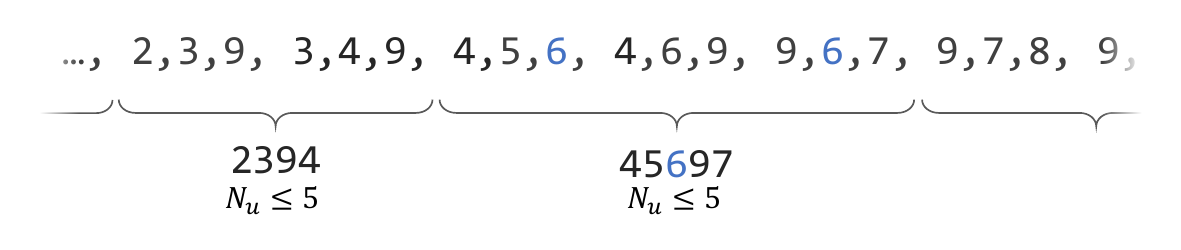}
		\vspace{-18pt}
		\caption{dynamic batching}
		\label{fig:dynamic_batching}
	\end{subfigure}
	\caption{
		The input can either be divided into (a) batches of a constant size $N_p$ (static batching) or (b) batches of a variable size chosen such that the number of unique vertices stays below a threshold $N_u$ (dynamic batching).
	}
	\label{fig:batching}
\end{figure}

An obvious challenge in the parallel generation of this many-to-one mapping is that the input-to-output ratio is not known in advance. 
Ideally, we would like to choose a batch size such that the number of unique vertices equals the block size, and one thread can run exactly one instance of the vertex shader. 
If that is not the case, under-utilization will arise, as threads that receive no unique vertex to process will simply idle. 
With larger batch sizes, it may be possible to identify a larger number of duplicate indices, at the cost of requiring multiple rounds of shader invocations to finish a whole batch. 
These considerations lead to the proposal of two strategies outlined in \figref{fig:batching}: \emph{static batching} and \emph{dynamic batching}.


\subsection{Static batching}
\label{sec:static}
For static batching, each thread block simply fetches a fixed number of indices from the input buffer to process. 
As a guideline for efficient processing, we use a common multiple of the block size and the primitive size as batch size, \eg, for triangles and thread block size $32$, we could use any multiple of $3\cdot 32 = 96$.
Since the batch size is fixed, static batching requires no preprocessing of the index buffer and can be applied directly to the input of a streaming pipeline.

\paragraph*{Statically batched na\"ive}
As a baseline, we implement a na\"ive strategy that does not attempt any vertex reuse. Instead, every thread is directly assigned to a primitive, and invokes the vertex shader for all its indices. As thread blocks always fetch the same number of indices, the static batch size is implicitly given. Notice that, while this strategy leads to duplicate vertex shader execution, it avoids all communication overhead. Thus, for very simple vertex shaders, this na\"ive approach may in fact show very good performance.

\begin{figure}
	\centering
	\includegraphics[width=\linewidth]{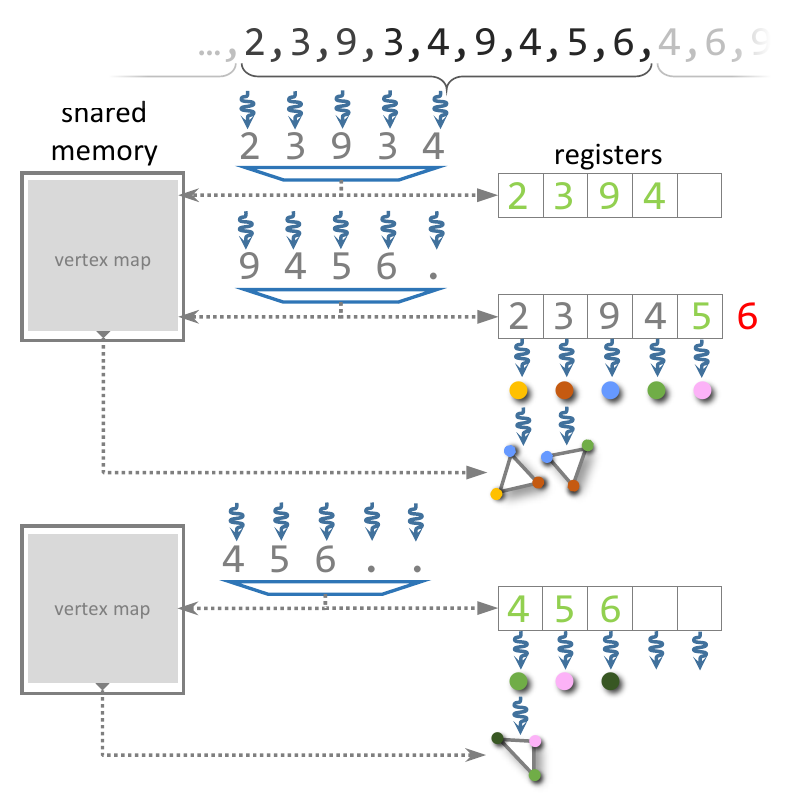}
	\caption{Statically-batched warp voting uses all threads in a warp (5 in this example) to load indices. We exploit warp voting and shuffle instructions to unify the indices and store the result in local registers. This process is repeated until all indices have been consumed, or all threads have acquired a unique index for processing in the vertex shader. As primitive size must be considered, early shading results might be discarded (\eg for index 5 above).}
	\label{fig:static_warp}
\end{figure}

\begin{algorithm}[h!]
	\DontPrintSemicolon
	\SetKw{Sync}{syncthreads}
	\SetKw{To}{to}
	\SetKw{Shared}{shared}
	\SetKw{MaxBatchSize}{BatchSize}
	\SetKw{BlockSize}{BlockSize}
	\SetKw{Sort}{BitonicSort}
	\SetKw{PrefixSum}{PrefixSum}
	\SetKw{VertexShader}{shade}
	\SetKw{Lid}{laneId}
	\SetKw{Output}{output}
	\SetKw{IIf}{if}
	\SetKw{And}{and}
	\SetKw{IElse}{else}
	\SetKw{Size}{size}
	\SetKw{Shfl}{shfl}
	\SetKw{Ballot}{ballot}
	\SetKw{LShift}{BitShift}
	\SetKw{Ffs}{ffs}
	\SetKw{Or}{or}
	\SetKw{Min}{min}

    
    \Shared map[ ]\;

    $cStart \leftarrow BatchBegin$\;
    \While{$cStart$ $<$ $BatchEnd$}{
        $fill$  $\leftarrow$ $0$, 
        $done$ $\leftarrow$ $0$, 
		$my\_id$ $\leftarrow$ $-1$,
        $offset$ $\leftarrow$ $cStart$\;
        \While{$offset < BatchEnd$ \And $fill < WarpSize$}{
        	$incoming \leftarrow -1$, 
			$outgoing \leftarrow -1$\;
            \If{$offset + \Lid < BatchEnd$}{$incoming \leftarrow$ indexBuffer[$offset + \Lid$]\;}
            
            \For{$i \in WarpSize$}{
                $current \leftarrow$ \Shfl($incoming$, $i$)\;
                $match \leftarrow$  \Ballot($current = my\_id$)\;
                \If{$match = 0$}{
                    \If{$fill = \Lid$}{
						$my\_id$ = $current$\;
						}
					$match \leftarrow$ \LShift($1$, $fill$)\;
					$fill \leftarrow fill + 1$\;
                }
                \If{$i = \Lid$}{
                    $outgoing \leftarrow match$\;
                }	
            }
            
            map[$done+\Lid$] $\leftarrow$ \Ffs($outgoing$)-1\;
            $firstmask \leftarrow$ \Ballot($outgoing = 0$ \Or $incoming = -1$)\;
            $additional$ $\leftarrow$ \Min($WarpSize$, \Ffs($firstmask$) \;
			$done \leftarrow done + additional$	
        
            $offset \leftarrow offset + WarpSize$\;
        }
        $triangles \leftarrow \lfloor done/3 \rfloor$\;
        
        \If{$\Lid < fill$}{
            $v \leftarrow$ \VertexShader(vertexBuffer[$my\_id$])\;
        }
        $v0 \leftarrow$ \Shfl($v$, map[$3\cdot\Lid$])\;
        $v1 \leftarrow$ \Shfl($v$, map[$3\cdot\Lid+1$])\;
        $v2 \leftarrow$ \Shfl($v$, map[$3\cdot\Lid+2$])\;
        \If{\Lid $<$ $triangles$}{
            \Output($v0$,$v1$,$v2$)\;
        }
        $cStart \leftarrow 3 \cdot triangles$\;
    }
	\caption{Statically batched warp voting.}
	\label{alg:warpvote}
\end{algorithm}

\paragraph*{Statically batched warp voting}
For this strategy, we aim to fill up warps with triangles so that every thread receives a unique vertex to work on, as outlined in \figref{fig:static_warp}.
For this purpose, we use fast, warp-level communication mechanisms, as detailed in Algorithm~\ref{alg:warpvote}.
Every thread first loads an index from the buffer and subsequently publishes it via register shuffle instructions to all other threads in the warp. 
Each thread then informs its peers via warp voting whether a duplicate index has been found. 
We track the number of unique indices observed so far and assign each new index to the available thread with the lowest ID.
We also maintain an inverse lookup-table in shared memory for fast reassembly after shading.

We keep fetching indices until either all threads were assigned a unique vertex, or the batch boundary is hit. 
Next, all identified unique vertices are shaded and output assembly is carried out.
This process is repeated iteratively, until all indices in the batch have been processed.
Note that starting a new iteration can lead to duplicate shader invocation inside a batch, since shading results are not carried over from the previous iteration.
To distribute the shaded vertices within the warp, we again use shuffle instructions.

\subsection{Dynamic batching}
We assess the potential for optimizing vertex processing by allowing for a fast, low-impact preprocessing step to retrieve analytical data from the submitted index buffer.
Specifically, we investigate the performance of several dynamic batching strategies, which rely on a load-time analysis of the input to derive optimal batch sizes. 
This routine splits the buffer into batches of variable length, with the goal of maximizing thread occupancy at runtime for the loading and processing of vertices.

To achieve this, we define $N$ to be a multiple of the block size and scan the triangles in the index buffer front to back, counting unique indices until we reach $N$, or a maximum allowed number of primitives has been added to the batch. 
As soon as either of these conditions is met, we start a new batch and continue scanning the index buffer until all indices have been assigned to their respective batches.
The batch starting positions, stored in an auxiliary buffer, allow us to feed a close-to-ideal amount of data to each thread block.

\begin{figure}
	\centering
	\includegraphics[width=\linewidth]{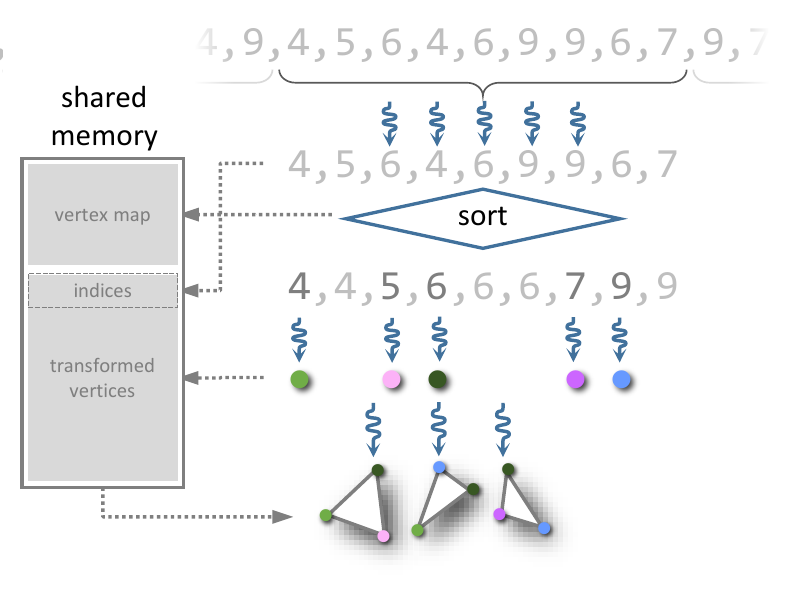}
	\caption{Dynamically-batched sorting brings indices into a monotonic order using radix sort. This allows to efficiently determine unique indices. Computing the prefix sum yields offsets for the individual indices to store shading results at. An inverse mapping is used to assemble the output primitives.}
	\label{fig:dynamic_sorting}
\end{figure}

\begin{algorithm}[h!]
	\DontPrintSemicolon
	\SetKw{Sync}{syncthreads}
	\SetKw{To}{to}
	\SetKw{Shared}{shared}
	\SetKw{MaxBatchSize}{BatchSize}
	\SetKw{BlockSize}{BlockSize}
	\SetKw{Sort}{RadixSort}
	\SetKw{PrefixSum}{PrefixSum}
	\SetKw{VertexShader}{shade}
	\SetKw{Tid}{tId}
	\SetKw{Output}{output}
	\SetKw{IIf}{if}
	\SetKw{IElse}{else}
	\SetKw{Size}{size}

    
    \Shared ids[ ],  linIds[ ], map[ ], marks[ ], uniqueIds[ ], v[ ]\;

	\ForPar{$i \in \Size(Batch)$}{
	    ids[$i$] $\leftarrow$ indexBuffer[$BatchBegin + i$]\;
	    linIds[$i$] $\leftarrow$ $i$\;
	}
    \Sort(ids, linIds)\;
    \ForPar{$i \in \Size(Batch)$}{
        marks[$i$] $\leftarrow$ $1$ \IIf ids[$i$] $\neq$ ids[$i+1$] \IElse $0$ \; 
	}
	$numVertices \leftarrow$\PrefixSum(marks)\;
	\ForPar{$i \in \Size(Batch)$}{
	    map[linIds[$i$]] $\leftarrow$ marks[$i$]\;
	    uniqueIds[marks[$i$]] $\leftarrow$ ids[$i$]\;
	    
	    }
    \ForPar{$j \in numVertices$}{
	    v[$j$] $\leftarrow$ \VertexShader(vertexBuffer[uniqueIds[$j$]])\;
	}
	\ForPar{$i \in \Size(Batch)/3$}{
       \Output(v[map[$3i$], v[map[$3i+1$], v[map[$3i+2$]])\;
	}
	
	\caption{Dynamically batched sorting.}
	\label{alg:sorting}
\end{algorithm}

Note that we do not require the buffer to forward information about the unique vertices, and leave their identification to be conducted by threads at runtime.
Hence, no information other than the splitting of the index buffer into optimally processable portions is output at this point.
Therefore, we can abstract our preprocessing procedure to an elaborate work scheduling routine, that could very well be realized by dedicated hardware.
We propose three distinct strategies for the dyanimc approach: \term{dynamically batched sorting}, \term{dynamically batched hashing}, \term{dynamically batched parallel hashing}.

\paragraph*{Dynamically batched sorting.}
One straight-forward way to determines the assignment between threads and unique vertices is to use parallel sorting. 
We load and sort a full batch of indices in shared memory, and run a prefix sum over the sorted sequence to determine all unique vertex indices. 
The original position in the batch is carried along during sorting, to be used as an inverted lookup-table for assembly as outlined in Algorithm~\ref{alg:sorting} and \figref{fig:dynamic_sorting}.

\begin{figure}
	\centering
	\includegraphics[width=\linewidth]{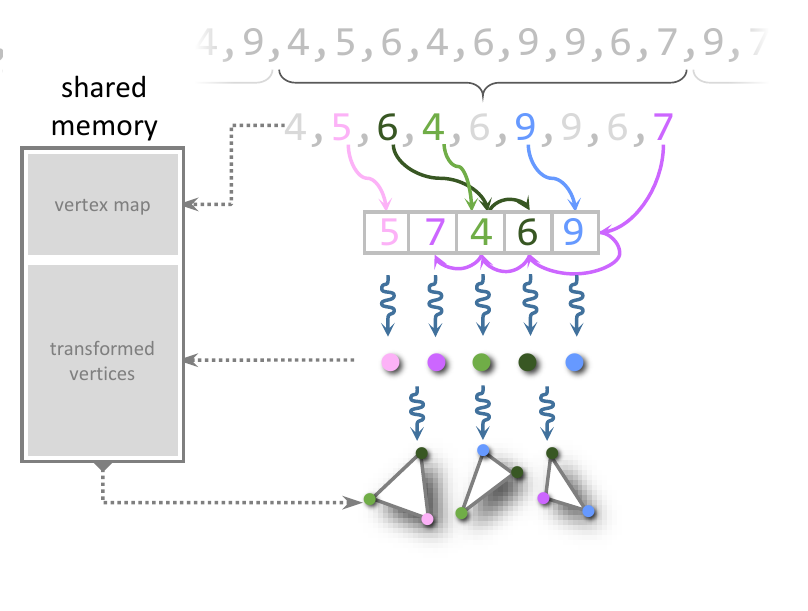}
	\caption{Dynamically-batched hashing uses a hash map that can hold one entry for each thread. Filling up the hash map removes duplicates in the data and each thread can execute the vertex shader on one unique vertex.}
	\label{fig:dynamic_hashing}
\end{figure}

\noindent 
\paragraph*{Dynamically batched hashing.}
In this strategy, we employ a hash map in shared memory to remove duplicate vertex indices. 
As hash function, we use multiplicative hashing with linear probing. 
We choose the size of the hash map to match the thread block size. 
Ideally, the hash map is fully filled after loading a batch due to the size restrictions applied in our preprocessing step.
Consequently, filling the hash map allows us to uniquely assign vertices to threads. 
Upon entering an index into the hash map, the loading thread records the value of the hash function, to be identify the required vertex after shading.
Since the hash map is filled in parallel, we use atomic operations for insertion, as outlined in Algorithm~\ref{alg:hashing} and \figref{fig:dynamic_hashing}.

\begin{algorithm}
	\DontPrintSemicolon
	\SetKw{Sync}{syncthreads}
	\SetKw{To}{to}
	\SetKw{Shared}{shared}
	\SetKw{MaxBatchSize}{BatchSize}
	\SetKw{BlockSize}{BlockSize}
	\SetKw{Sort}{BitonicSort}
	\SetKw{PrefixSum}{PrefixSum}
	\SetKw{VertexShader}{shade}
	\SetKw{Tid}{tId}
	\SetKw{Output}{output}
	\SetKw{IIf}{if}
	\SetKw{IElse}{else}
	\SetKw{Size}{size}
	\SetKw{Cas}{atomicCAS}
	\SetKw{Hash}{hash}
	\SetKw{Prob}{probing}
	\SetKw{Or}{or}

    
    \Shared hashtable[ ], map[ ], v[ ]\;

	\ForPar{$i \in \MaxBatchSize$}{
	    hashtable[$i$] $\leftarrow$ -1\;
	}
    \ForPar{$i \in \Size(Batch)$}{
        $id \leftarrow$ indexBuffer[$BatchBegin + i$]\;
        $p \leftarrow$ \Hash($id$)\;
        \While{not inserted}{
            $prev \leftarrow$ \Cas(hashtable[$i$], $-1$, $id$ )\;
            \uIf{$prev - 1$ \Or $prev = id$ }{
                $loc \leftarrow p$\;
            }
            \Else{
                $p \leftarrow$ \Prob(p)\;
            }
        }
        map[$i$] = $loc$;
	}
	
	\ForPar{$j \in \MaxBatchSize$}{
	    \If{hashtable[$j$] $\neq -1$}{
	        v[$j$] $\leftarrow$ \VertexShader(vertexBuffer[hashtable[$j$]])\;
	    }
	}
	\ForPar{$i \in \Size(Batch)/3$}{
       \Output(v[map[$3i$], v[map[$3i+1$], v[map[$3i+2$]])\;
	}    
	
	\caption{Dynamically batched hashing.}
	\label{alg:hashing}
\end{algorithm}

\noindent \paragraph*{Dynamically batched parallel hashing.}
One issue with the simple hashing approach above is, that a fully occupied hash map will likely lead to excessive linear probing. 
In some cases, this may lead to pathological warp divergence, as a single thread repeatedly tries to find the last free entry, and the remaining peers in the warp have to join in the effort. 
As a remedy, we propose to perform hashing as a two-tiered approach. 
First, every thread executes up to a fixed number of linear probing attempts. 
Second, all threads within a warp collaboratively collaborate to find available spots until all indices have been inserted. 
This fast-path/slow-path strategy effectively repurposes otherwise idle threads in order to speed up the search for free spots. 
Coordination within a warp can be realized through efficient register shuffle and warp voting.



\begin{figure*}
	\begin{minipage}{\linewidth}
		\begin{subfigure}{2ex}
			\rotatebox[origin=c]{90}{Original}
		\end{subfigure}\hfill%
		\begin{subfigure}{0.32\linewidth}
			\centering
			\includegraphics[width=\linewidth]{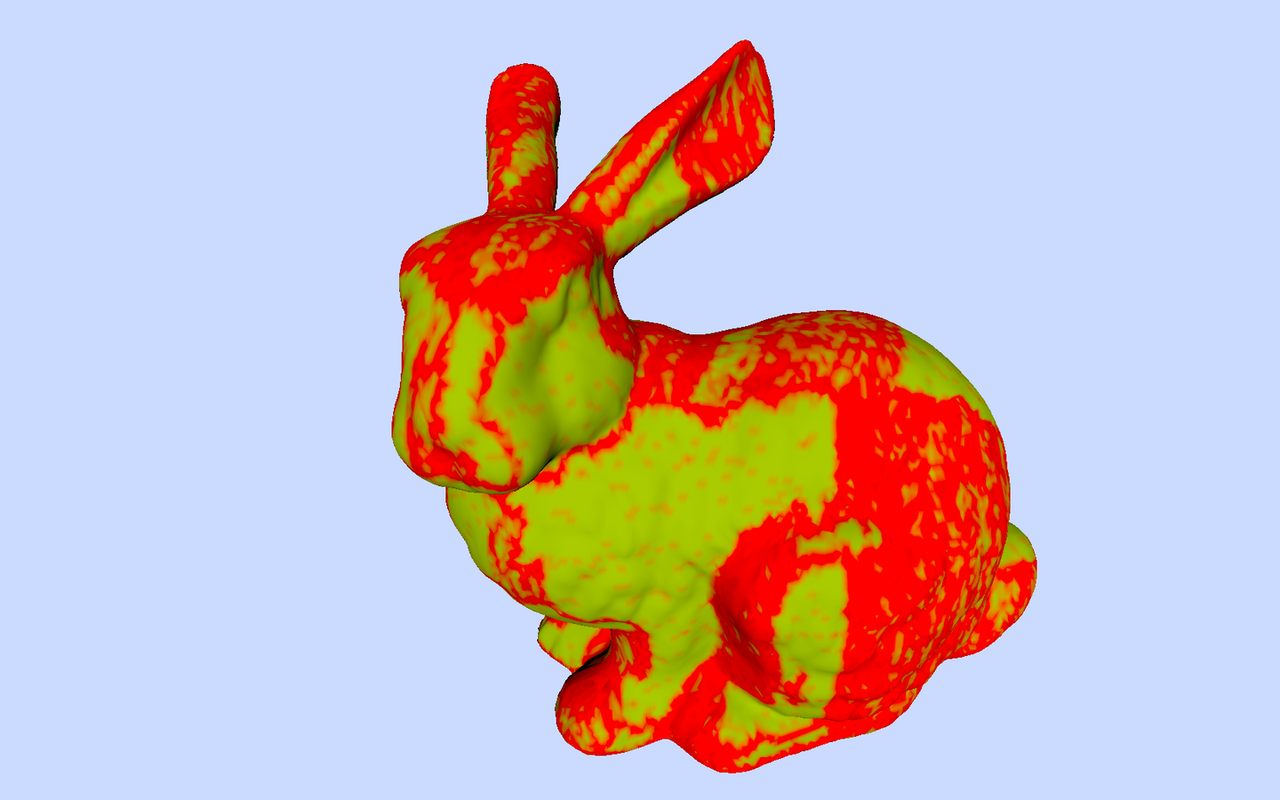}
		\end{subfigure}\hfill%
		\begin{subfigure}{0.32\linewidth}
			\centering
			\includegraphics[width=\linewidth]{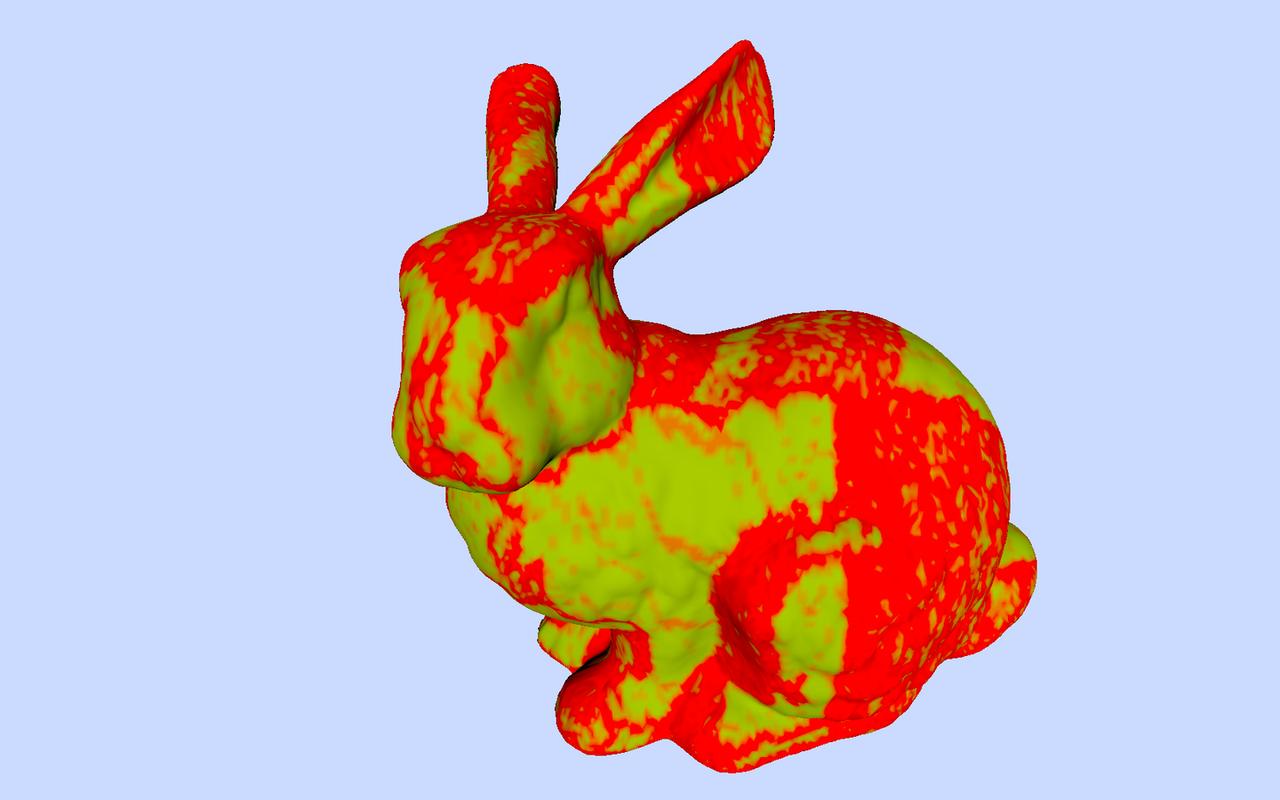}
		\end{subfigure}\hfill%
		\begin{subfigure}{0.32\linewidth}
			\centering
			\includegraphics[width=\linewidth]{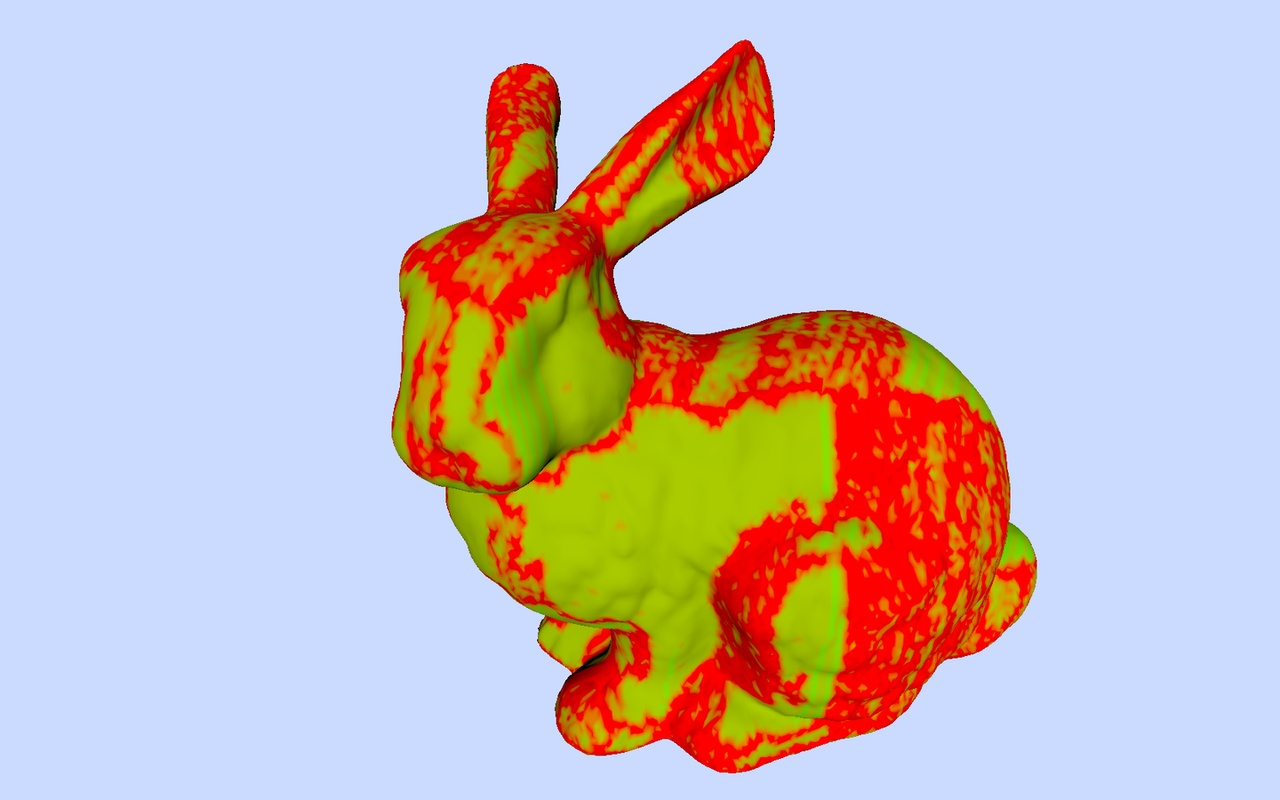}
		\end{subfigure}
	\end{minipage}\vspace{0.5ex}
	\begin{minipage}{\linewidth}
		\begin{subfigure}{2ex}
			\rotatebox[origin=c]{90}{TomF}
		\end{subfigure}\hfill%
		\begin{subfigure}{0.32\linewidth}
			\centering
			\includegraphics[width=\linewidth]{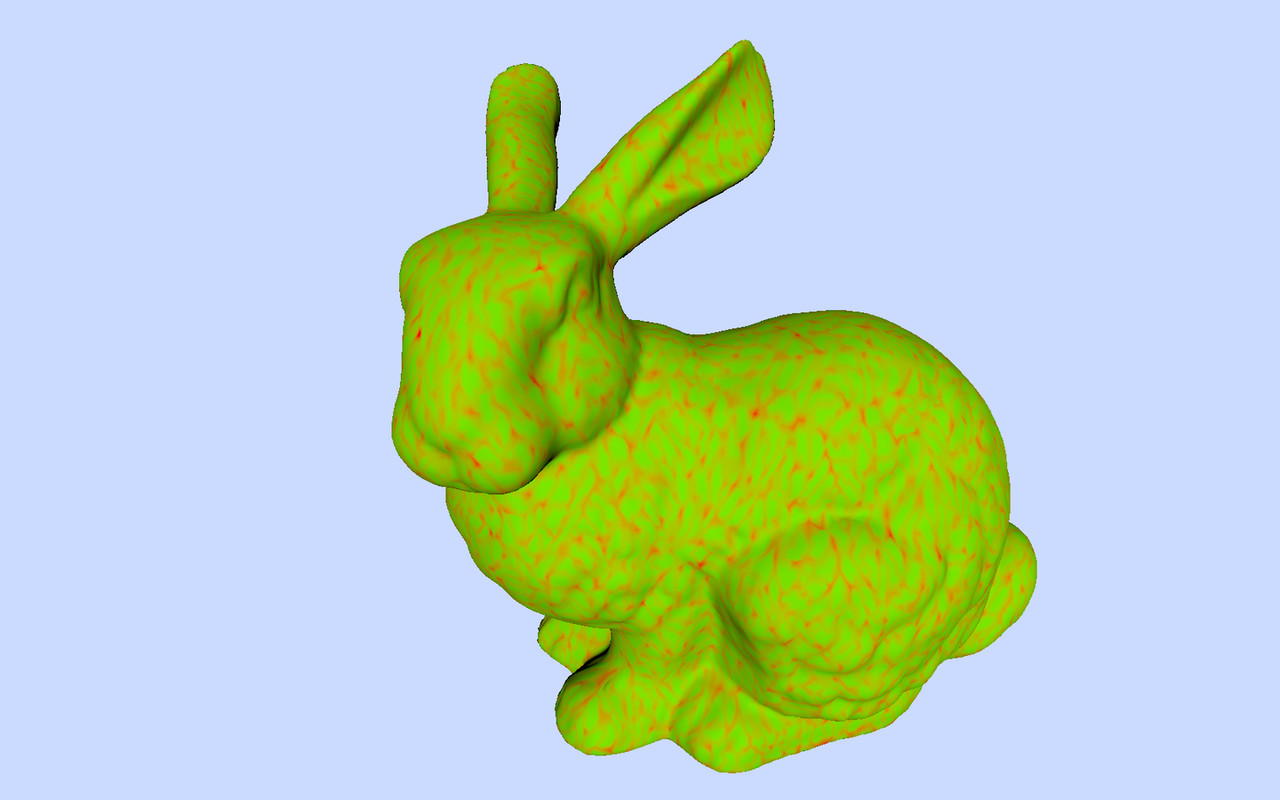}
			\caption{OpenGL}
		\end{subfigure}\hfill%
		\begin{subfigure}{0.32\linewidth}
			\centering
			\includegraphics[width=\linewidth]{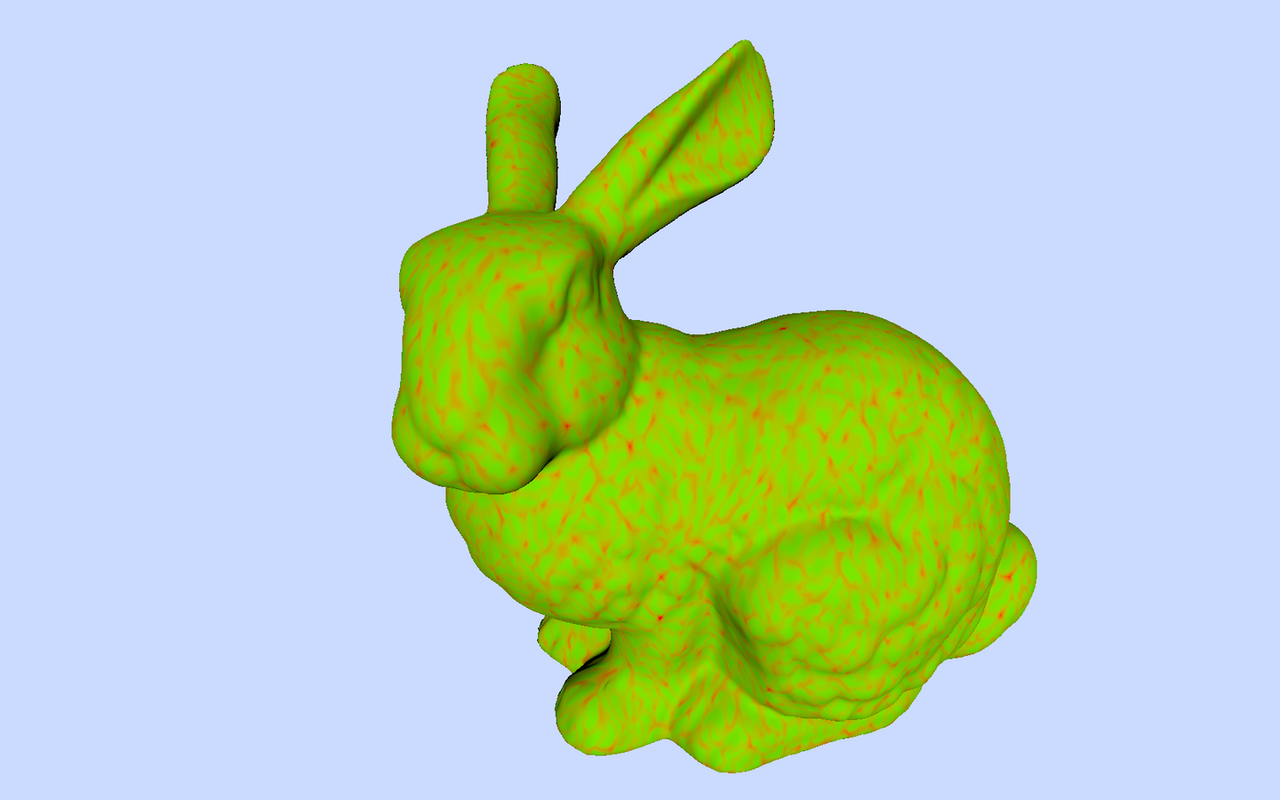}
			\caption{Static batching}
		\end{subfigure}\hfill%
		\begin{subfigure}{0.32\linewidth}
			\centering
			\includegraphics[width=\linewidth]{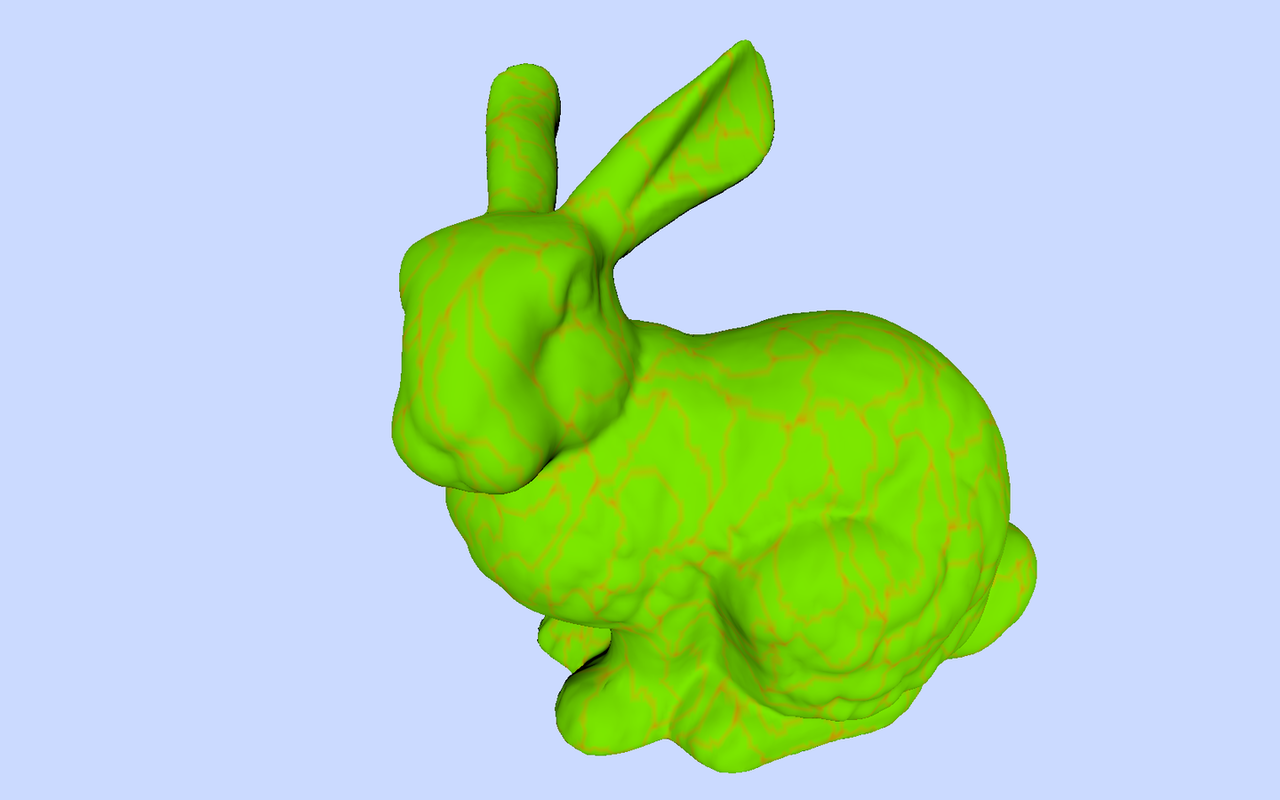}
			\caption{Dynamic batching}
		\end{subfigure}
	\end{minipage}
	\caption{
		Vertex reuse visualization for the Stanford Bunny model: green indicates a single shader invocation, red indicates six shader calls. Compared to the original input, preprocessing the model with TomF reduces arbitrary discontinuities and enables better overall potential for vertex reuse for both OpenGL and software techniques. While dynamic batching shows higher reuse, static warp voting seems to be closer to actual OpenGL behavior on a GTX 1080Ti.
	}
	\label{fig:reusevis_bunny}
\end{figure*}

\section{Evaluation}

For performance evaluation, we use a set of commonly processed models, as well as content captured from five recent video games and an NVIDIA technical demo: Age of Mythology (abbreviated am), Assassin's Creed: Black Flag (as), Deus Ex: Human Revolution (dx), Stone Giant animation (sg), Total War: Shogun 2 (sh),  Rise of the Tomb Raider (tr), and The Witcher 3 (tw). 
A representative rendering from our 19 different scenes is shown in Figure~\ref{fig:game}. 

\begin{figure}
	\centering
	\includegraphics[width=0.85\linewidth]{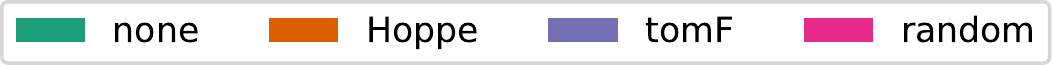}
	\begin{subfigure}{0.33\linewidth}
		\includegraphics[width=\linewidth]{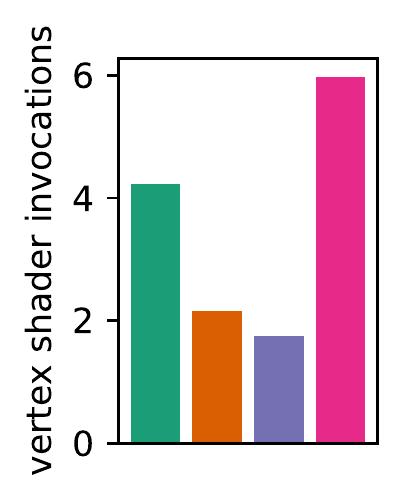}
		\caption{GL}
	\end{subfigure}%
	\begin{subfigure}{0.33\linewidth}
		\includegraphics[width=\linewidth]{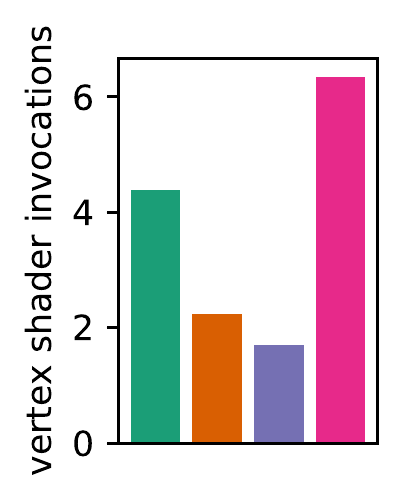}
		\caption{warp}
	\end{subfigure}
	\begin{subfigure}{0.33\linewidth}
		\includegraphics[width=\linewidth]{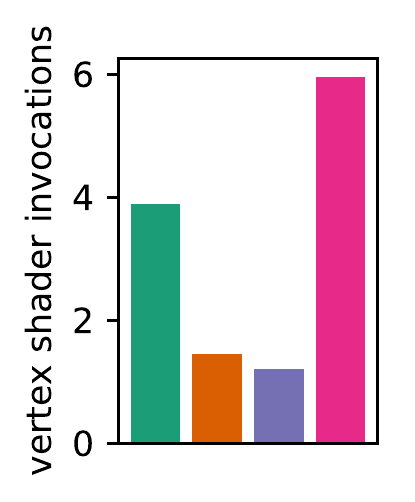}
		\caption{dyn.}
	\end{subfigure}
	\caption{Preprocessing the Stanford Bunny for vertex locality significantly improves its reuse potential. In OpenGL, every vertex is shaded four times in the original mesh; a random organization leads to nearly six invocations; preprocessing with TomF~\cite{forsyth2006linear} or Hoppe~\cite{Hoppe:1999:OML:311535.311565} reduces the number of shader calls. Our approaches (warp, dyn.) mirror this behavior.}
	\label{fig:bunnyshadercall}
\end{figure}

Obviously, evaluating the effectiveness of different vertex reuse methods is meaningful only if the processed data actually allows for detecting reuse. 
As this is usually not considered in the generation, formatting or conversion of mesh data, the order in which vertices are referenced in input models can be at best arbitrary, or, at worst, biased. 
Popular mesh processing algorithms have been presented previously, with the aim of reordering indices in a given mesh to increase vertex locality. 
As shown by Figures \ref{fig:reusevis_bunny} and \ref{fig:bunnyshadercall}, applying such algorithms to popular models can remove unusual discontinuities, and significantly improve reuse potential in the OpenGL streaming pipeline.
Incidently, this is also true for our own techniques, which appear to exhibit similar behavior to OpenGL. 
In order to generate a fair ordering  and enable vertex reuse even in unstructured models, we preprocess all meshes with the optimization algorithm by \citeauthor{forsyth2006linear} (\citeyear{forsyth2006linear}). 

For comparison, we simulate a parallel caching scheme with different cache sizes and compare performance to our static and dynamic batching approaches. 
Furthermore, we investigate performance of our techniques when using them in a software streaming rendering pipeline and compare to a non-streaming, multi-kernel setup.

\subsection{Caching vs batching and OpenGL}
We first determine the ideal reuse rate as the ratio of duplicate vertex indices over the total length of the index buffer.
Theoretically, a very large, global post-transform cache with instant reusability could yield the reported ideal figures.
Unfortunately, such a global vertex cache does not seem practical for massively parallel devices, like modern GPUs.
Storing and retrieving data from device-wide caches introduces an order of magnitude higher latency than caches on multi-processors, effectively reducing performance significantly.
Furthermore, as cache entries can only be generated after vertex shader execution, all threads that concurrently receive the same non-cached index to execute,  will not be captured by the cache.
On a massively parallel device, like the GPU, tens of thousands of threads may execute in parallel, precluding an immense vertex reuse potential of being utilized.
The problem gets even worse considering the long latency of a device-wide cache, \eg, even if one thread completes a vertex shader invocation and another threads queries the cache for that vertex just after, it may still result in a cache miss, as the latency for storing the shading result has not yet elapsed. 

In order to compare to a more realistic cache-based approach, we simulate a per-multiprocessor cache setup.
Based on the design of our main evaluation GPU, the NVIDIA GTX 1080Ti, we simulate 28 multi-processors, where each multi-processor runs 1024 vertex shader invocations in parallel and stores the results in its dedicated least-recently-used (LRU) cache of size 16, 32, or 64KB.
As mentioned before, a cache hit cannot occur for duplicate vertices being concurrently processed, but only if the result was already produced in an earlier shading cycle.
%
For our statically batched warp voting, we use a batch size of 96, which fits the warp size of 32.
For all dynamic batching approaches, we use a maximum batch size of 1023 indices and 256 unique vertices, and assign 256 threads to each batch. 
Test scene statistics and achieved reuse for each technique are listed in Table~\ref{tab:statistics}.

\def\rtbs{\hspace{3.4pt}}
\begin{table}[h!]
	\caption{Scene statistics: vertices, triangles, and average vertex reuse (equivalent to cache hit rates) in an ideal case, for a per-multiprocessor cache, warp voting, and dynamic batching.}
	\centering
	\begin{tabular}{@{\rtbs}lc@{\rtbs}@{\rtbs}cc@{\rtbs}@{\rtbs}c@{\rtbs}@{\rtbs}c@{\rtbs}@{\rtbs}c@{\rtbs}@{\rtbs}c@{\rtbs}@{\rtbs}c@{\rtbs}}
		\toprule
				& & & & \multicolumn{3}{c}{Parallel Cache} & \multicolumn{2}{c}{Ours} \\
				& vert 	& tris 	& ideal & 16KB 	& 32KB 	& 64KB	& warp 	& dyn. \\
		\midrule
		bunny 	& 34k  	& 70k  	& .832 	& .056	& .060 	& .067			& .714 	& .799  \\
		sphere 	& 40k  	& 82k  	& .833 	& .063	& .065 	& .069 			& .712 	& .795  \\
		tree   	& 492k 	& 239k 	& .314 	& .001	& .001 	& .001 			& .314 	& .313  \\
		buddha 	& 544k 	& 1.1M 	& .833 	& .050	& .048 	& .049  		& .725 	& .804  \\
		dragon 	& 3.6M 	& 7.2M 	& .833 	& .059	& .059 	& .059 			& .713 	& .795  \\
		am02   	& 3k 	& 6k 	& .801 	& .042	& .057 	& .075 			& .709 	& .779  \\
		am03   	& 2k 	& 4k 	& .839 	& .090	& .108 	& .108 			& .735 	& .807  \\
		as01   	& 108k 	& 183k 	& .803 	& .034	& .035 	& .035 			& .719 	& .788  \\
		as04   	& 598k 	& 538k 	& .629 	& .017	& .018 	& .018			& .575 	& .620  \\
		dx29   	& 25k 	& 42k 	& .796 	& .034	& .035 	& .037			& .719 	& .782  \\
		dx33   	& 37k 	& 60k 	& .795 	& .028	& .029 	& .029			& .718 	& .784  \\
		sg14   	& 135k 	& 254k 	& .822 	& .043	& .044 	& .044			& .726 	& .805  \\
		sg16   	& 38k 	& 69k 	& .813 	& .047	& .048 	& .048			& .720 	& .796  \\		
		sh11   	& 812k 	& 1.1M 	& .754 	& .025	& .026 	& .026			& .693 	& .744  \\
		sh21   	& 521k 	& 701k 	& .751 	& .029	& .029 	& .029			& .681 	& .737  \\
		tr04   	& 191k 	& 283k 	& .775 	& .033	& .034 	& .034			& .708	& .763  \\
		tr09   	& 78k 	& 118k 	& .780 	& .031	& .031 	& .031			& .705 	& .769  \\
		tw03   	& 268k 	& 487k 	& .816 	& .051	& .052 	& .053			& .709 	& .787  \\
		tw30   	& 695k 	& 565k 	& .589 	& .020	& .020 	& .020			& .532 	& .579  \\
		\bottomrule
	\end{tabular}
	\label{tab:statistics}
\end{table}

The recorded reuse rates of these cache-based techniques is always below $\frac{1}{5}$ of the ideal reuse rate. 
In contrast, our dynamic batching usually falls within 5\% of the ideal reuse, and statically batched warp voting within 10--20\%. 
In general,  both static and dynamic batching seem to be highly effective for all kinds of scenes.

The sum of these observations strengthens our assumption that the herein presented approaches may be better suited for modern architectures than a post transformation cache.
Furthermore, our tests with implementing a post-transform cache, quickly lead us to believe that there is no efficient way of implement such a cache in software.
Thus, we solely focus on the presented batch-based approaches throughout the evaluation of use-case scenarios.

\subsection{Rasterization}

The major motivation and use case for vertex reuse is the geometry processing stage of a real-time 3D rendering pipeline.
To test our batch-based reuse techniques, we have implemented a configurable geometry stage in CUDA, that can be included into a streaming pipeline design.
The geometry processing stage is simply given an input stream of indices and a vertex buffer.
Based on the respective batching approach, indices are fetched from the index buffer and vertex reuse is evaluated.
As a final step, one thread per triangle is used to write the output primitive with its vertices into a queue.
This output queue could---when integrated into a full streaming pipeline---be consumed by the next stage in the rendering pipeline.
For a traditional real-time rendering pipeline, this queue would form the natural point for a sort middle approach.

The runtime results for the vertex processing for selected tested techniques on an NVIDIA GTX 1080 Ti are shown in Table~\ref{tab:results}. 
To simulate different vertex shader loads, we used a simple matrix multiplication (simple), a load of 256, 512 and 1024 cycles; for comparison, a cached access to L1 takes about 20 cycles, to L2 about 200. 
We also include a non-streaming, multi-staged processing implementation for reference.
With this approach, all vertices in the vertex buffer are processed only once by separate kernel.
The output vertex data can then be directly loaded from global memory in a separate kernel for assembling the output primitives.
This approach is employed, \eg, by \citeauthor{Laine:2011:HSR} (\citeyear{Laine:2011:HSR}) for rasterization of 3D scenes.
Since vertices need to be shaded exactly once, this technique can achieve ideal reuse, but only at the cost of sacrificing the advantages of a streaming architecture.
For instance, unreferenced vertices are shaded in this approach regardless of their relevance to the scene.
This can exact a significant performance penatly if, \eg, one large vertex buffer is used in combination with multiple index buffers to draw relevant portions on demand.
Although this is not the case in our test scenes, our vertex reuse techniques can even outperform multi-staged geometry processing on several accounts.
The full data set we produced for evaluating reuse in our software renderer can be found in the accompanying supplemental material for this paper. 

As can be seen, for a very simple vertex shader, the na\"ive, no-reuse approach is the fastest, as it has no communication overhead. 
However, warp voting and sorting are on average only about $1.5\times$ slower, and both hashing approaches are about $2.0\times$ behind. The results indicates to us that, among the techniques capable of vertex reuse, warp voting has the lowest overhead.

As the vertex shader load increases, the na\"ive approach quickly falls behind, indicating that the proposed approaches can efficiently detect vertex reuse. 
For a 256 cycle load, warp-voting achieves the best performance, followed by parallel-hashing and hashing. 
For this load, the lower overhead of warp-voting still outweighs its lower vertex reuse rate. 
However, for larger loads the two hashing approaches catch up, and performance is overall tied between the our three techniques, while sorting eventually trails behind. 
We attribute the high performance of both hashing approaches and their marginal difference to the efficient implementation of shared memory atomics on recent GPUs.

	\begin{table}
	\caption{Processing times achieved with different vertex reuse techniques for rendering of geometry on a GTX 1080Ti. We also include a non-streaming, multi-kernel technique (\emph{multi}) for comparison.}
	\label{tab:results}
		\centering
\def\rtbs{\hspace{3pt}}
\begin{tabular}{@{\rtbs}c@{\rtbs}l@{\rtbs}c@{\rtbs}c@{\rtbs}c@{\rtbs}c@{\rtbs}c@{\rtbs}c@{\rtbs}c@{\rtbs}c@{\rtbs}c@{\rtbs}c@{\rtbs}}\toprule
 &  & sph & tree & dra & bud & as01 & dx33 & sg14 & sh11 & tr04 & tw03 \\
\midrule\multirow{6}{*}{\rotatebox[origin=c]{90}{simple}}
 & na\"ive & \textbf{0.07} & \textbf{0.34} & \textbf{4.59} & \textbf{0.81} & \textbf{0.14} & \textbf{0.05} & \textbf{0.19} & \textbf{0.83} & \textbf{0.21} & \textbf{0.35} \\
 & warp & 0.13 & 0.40 & 10.38 & 1.69 & 0.24 & 0.11 & 0.34 & 1.39 & 0.36 & 0.61 \\
 & hash & 0.16 & 0.42 & 12.89 & 2.05 & 0.27 & 0.12 & 0.46 & 1.47 & 0.40 & 0.74 \\
 & phash & 0.13 & 0.41 & 11.00 & 1.82 & 0.23 & 0.10 & 0.39 & 1.32 & 0.35 & 0.65 \\
 & sort & 0.20 & 0.51 & 6.36 & 1.21 & 0.31 & 0.13 & 0.41 & 1.18 & 0.34 & 0.61 \\
 & \textit{multi} & \textit{0.10} & \textit{0.38} & \textit{6.24} & \textit{1.02} & \textit{0.19} & \textit{0.08} & \textit{0.26} & \textit{1.09} & \textit{0.30} & \textit{0.47} \\
\midrule\multirow{6}{*}{\rotatebox[origin=c]{90}{256 cycles}}
 & na\"ive & 0.22 & 0.66 & 14.71 & 2.57 & 0.47 & 0.17 & 0.64 & 2.57 & 0.72 & 1.18 \\
 & warp & \textbf{0.13} & \textbf{0.33} & \textbf{5.38} & \textbf{0.93} & \textbf{0.19} & \textbf{0.11} & \textbf{0.26} & \textbf{0.93} & \textbf{0.27} & \textbf{0.43} \\
 & hash & 0.16 & 0.47 & 9.27 & 1.60 & 0.30 & 0.14 & 0.49 & 1.39 & 0.40 & 0.69 \\
 & phash & 0.15 & 0.42 & 8.42 & 1.59 & 0.26 & 0.12 & 0.41 & 1.28 & 0.33 & 0.66 \\
 & sort & 0.22 & 0.66 & 8.19 & 1.69 & 0.37 & 0.15 & 0.51 & 1.56 & 0.54 & 0.87 \\
 & \textit{multi} & \textit{0.11} & \textit{0.49} & \textit{6.62} & \textit{1.14} & \textit{0.23} & \textit{0.09} & \textit{0.31} & \textit{1.24} & \textit{0.36} & \textit{0.54} \\
\midrule\multirow{6}{*}{\rotatebox[origin=c]{90}{512 cycles}}
 & na\"ive & 0.39 & 1.10 & 25.22 & 4.55 & 0.84 & 0.29 & 1.13 & 4.45 & 1.27 & 2.10 \\
 & warp & \textbf{0.17} & \textbf{0.53} & \textbf{6.38} & \textbf{1.13} & \textbf{0.27} & \textbf{0.14} & \textbf{0.35} & \textbf{1.27} & 0.38 & \textbf{0.56} \\
 & hash & 0.19 & 0.62 & 11.20 & 1.78 & 0.32 & 0.16 & 0.51 & 1.40 & 0.44 & 0.79 \\
 & phash & 0.18 & 0.55 & 8.48 & 1.64 & 0.29 & 0.15 & 0.44 & 1.32 & \textbf{0.37} & 0.74 \\
 & sort & 0.26 & 0.83 & 9.57 & 1.94 & 0.41 & 0.16 & 0.53 & 1.89 & 0.56 & 0.98 \\
 & \textit{multi} & \textit{0.13} & \textit{0.68} & \textit{7.82} & \textit{1.36} & \textit{0.27} & \textit{0.11} & \textit{0.38} & \textit{1.57} & \textit{0.44} & \textit{0.65} \\
\midrule\multirow{6}{*}{\rotatebox[origin=c]{90}{1024 cycles}}
 & na\"ive & 0.71 & 2.02 & 46.88 & 8.09 & 1.56 & 0.53 & 2.09 & 7.43 & 2.25 & 3.75 \\
 & warp & 0.25 & 0.98 & 10.90 & \textbf{1.78} & 0.43 & 0.21 & \textbf{0.56} & 2.14 & 0.65 & 0.95 \\
 & hash & 0.24 & 1.00 & 10.91 & 1.98 & 0.42 & 0.22 & 0.63 & 1.98 & 0.58 & 0.90 \\
 & phash & \textbf{0.23} & \textbf{0.91} & \textbf{10.29} & 2.01 & \textbf{0.37} & \textbf{0.20} & 0.57 & \textbf{1.86} & \textbf{0.54} & \textbf{0.89} \\
 & sort & 0.31 & 1.25 & 12.45 & 2.34 & 0.49 & 0.21 & 0.64 & 2.53 & 0.73 & 1.17 \\
 & \textit{multi} & \textit{0.16} & \textit{1.08} & \textit{10.24} & \textit{1.80} & \textit{0.36} & \textit{0.14} & \textit{0.51} & \textit{2.16} & \textit{0.62} & \textit{0.87} \\
\bottomrule
\end{tabular}

	\end{table}

\begin{figure*}
	\centering
	\includegraphics[width=0.5\linewidth]{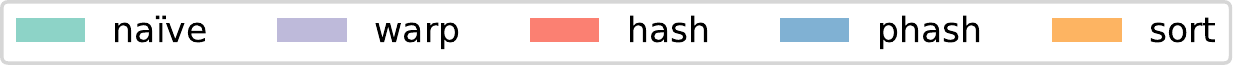}
	
	\begin{subfigure}{0.32\linewidth}
		\includegraphics[width=\linewidth]{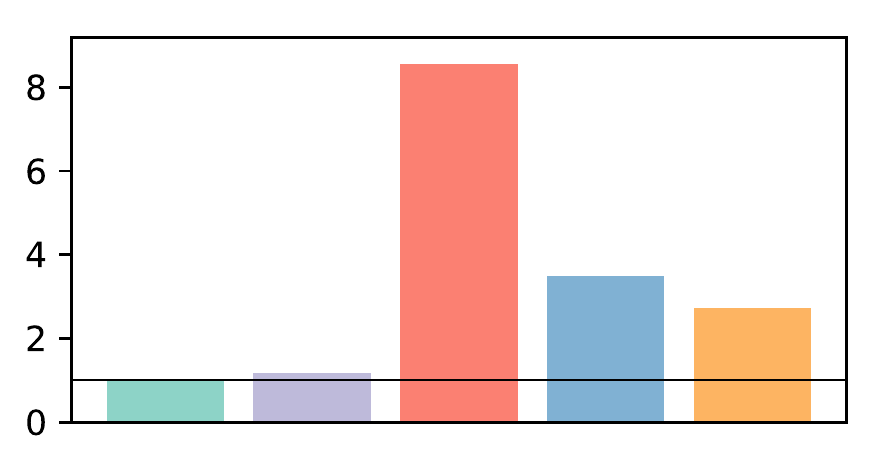}
		\caption{GTX 780 Ti simple}
	\end{subfigure}
	\begin{subfigure}{0.32\linewidth}
		\includegraphics[width=\linewidth]{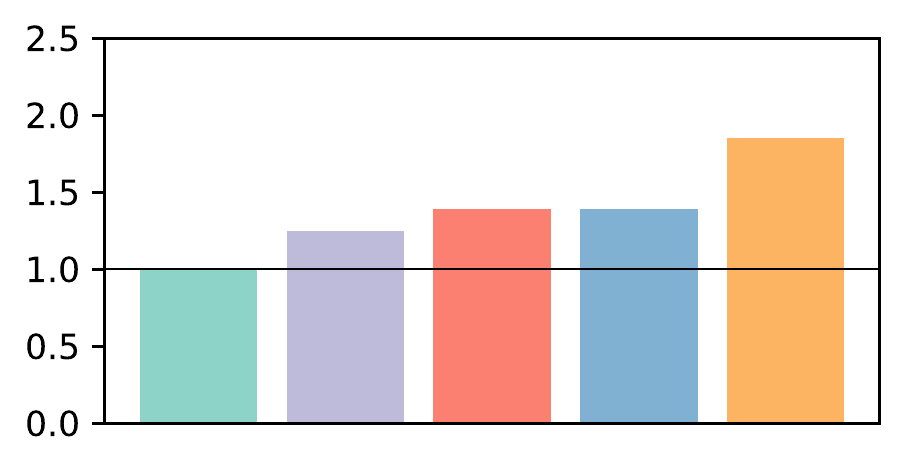}
		\caption{GTX 980 Ti simple}
	\end{subfigure}
	\begin{subfigure}{0.32\linewidth}
		\includegraphics[width=\linewidth]{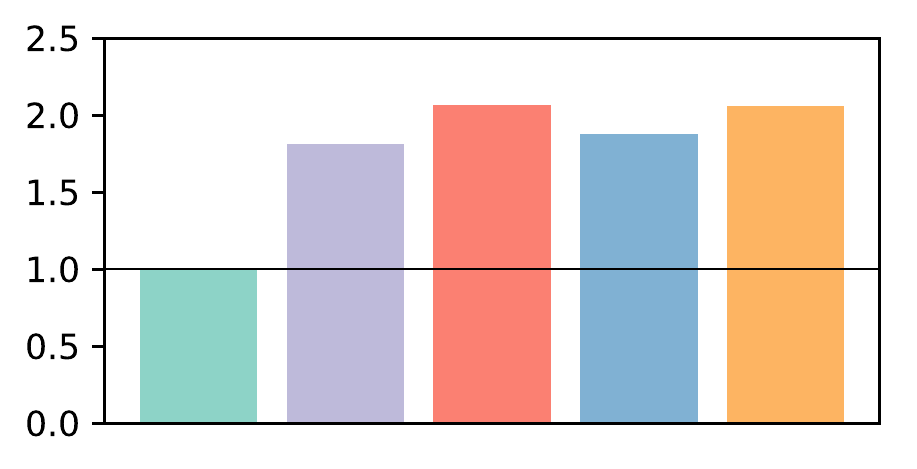}
		\caption{GTX 1080 Ti simple}
	\end{subfigure}
	
	\begin{subfigure}{0.32\linewidth}
		\includegraphics[width=\linewidth]{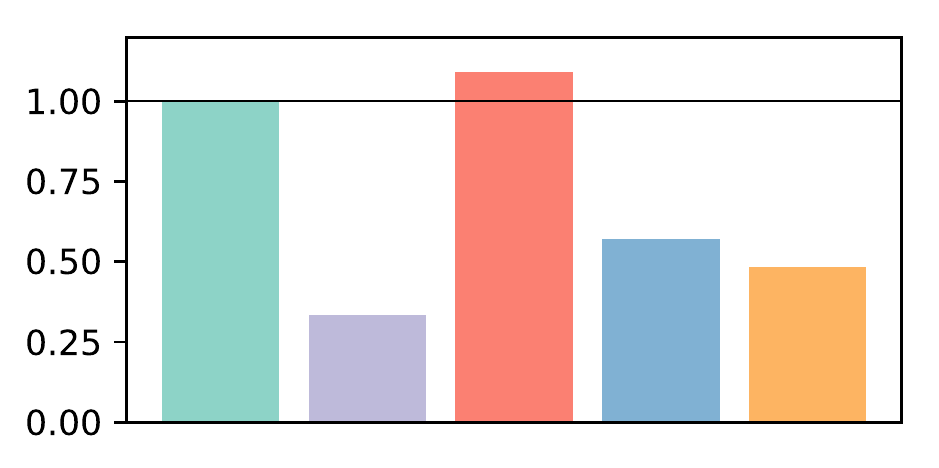}
		\caption{GTX 780 Ti 512 cycles}
	\end{subfigure}
	\begin{subfigure}{0.32\linewidth}
		\includegraphics[width=\linewidth]{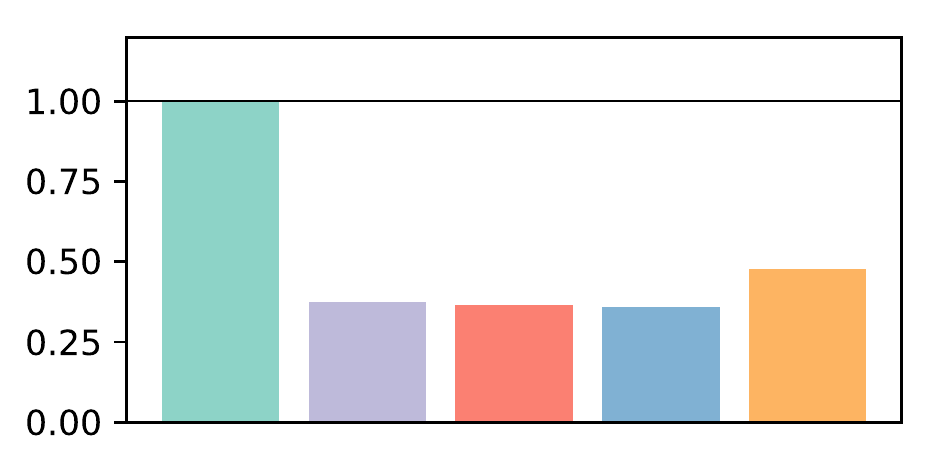}
		\caption{GTX 980 Ti 512 cycles}
	\end{subfigure}
	\begin{subfigure}{0.32\linewidth}
		\includegraphics[width=\linewidth]{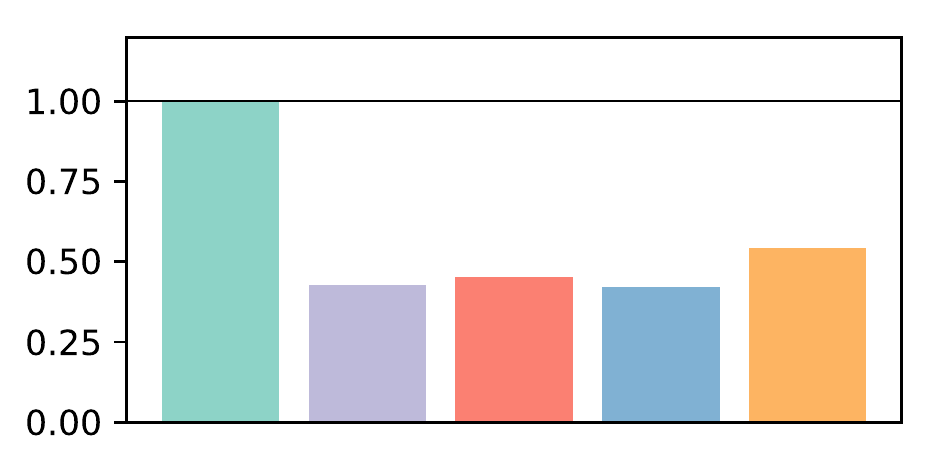}
		\caption{GTX 1080 Ti 512 cycles}
	\end{subfigure}
	\caption{Reported runtimes of the vertex processing stage in our software renderer, averaged over the entire test body (19 original scenes) and plotted relative to \id{na\"ive}. Different GPU architectures behave quite diversely: note the poor performance of \id{hash}, compared to our optimization \id{p.hash} on older architectures (GTX 780 Ti). Overall, statically batched warp voting seems to be the most reliable approach.}
	\label{fig:performance_avg}
\end{figure*}

To assess the performance of the proposed approach across multiple GPU generations, we also tested an NVIDIA GTX 780 Ti, 980 Ti and report the relative execution time compared to na\"ive, averaged over the entire test body in \figref{fig:performance_avg}. As can be seen, there is a significant difference across GPU generations, which is mostly due to more efficient shared memory operations.
While for a simple shader, all vertex reuse approaches reduce performance in comparison to na\"ive, the more complex shaders again benefit greatly from reuse. Although statically batched warp voting again slightly loses ground in comparison to the other approaches on the GTX 1080 Ti in the complex case, it outperforms the other approaches on average over all GPUs.
Additionally, statically batched warp voting does not require analysis of the index buffer and thus can be used in a full streaming approach. 

\section{Software Applications}
In addition to their respective impact to rendering performance, we also evaluate our techniques for its potential in context of general, software-based processing tasks that allow for reuse.
Specifically, we consider them for mesh subdivision and morphological transformation for inner and outer envelopes on 2-manifold models. 
Furthermore, we run a random walk simulation based on probabilistic input parameters. 
All applications were implemented in CUDA and executed on an NVIDIA GTX 1080Ti.

\subsection{Mesh Subdivision}
\begin{figure}
	\centering
	\begin{subfigure}{0.32\linewidth}
		\includegraphics[height=6.2cm]{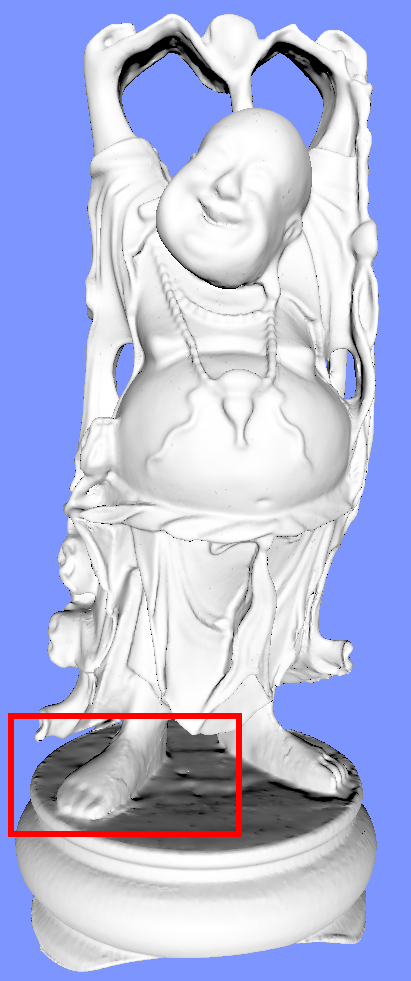}
		\caption{Buddha statue}
	\end{subfigure}%
	\hfill
	\begin{subfigure}{0.66\linewidth}
			\includegraphics[height=3.085cm]{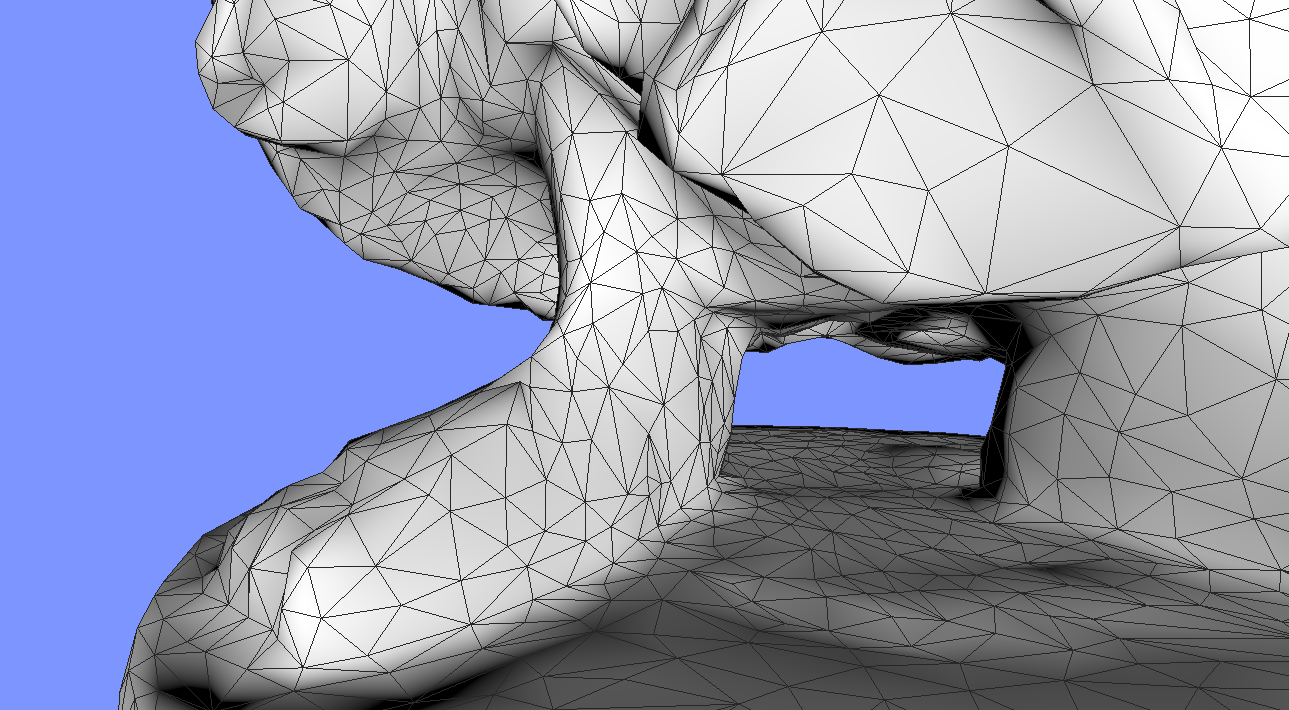}
			\includegraphics[height=3.085cm]{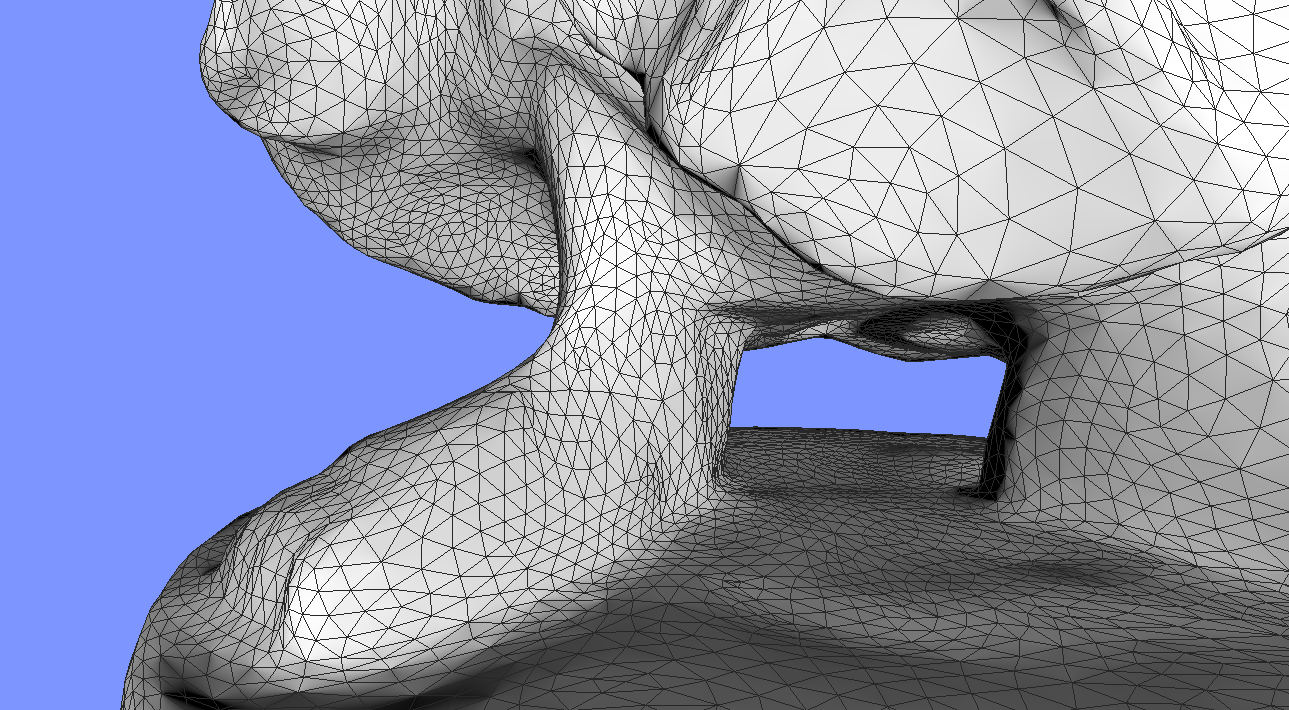}
			\caption{Wireframe close-up w/o and w/ subdivision}
	\end{subfigure}%
	\caption{Running Loop subdivision on a simplified \id{buddha} with vertex streaming. The smoother, subdivided output is shown in the lower right.}
	\label{fig:subdiv}
\end{figure}

Subdivision of low-detail meshes can be achieved by adding primitives to mesh, based on the adjacency information of its vertices.
The particular steps to take, as well as orientations of the newly introduced triangles are determined by analyzing their topological neighborhood.
An easily parallelizable subdivision algorithm has been presented in \cite{smooth-subdivision-surfaces-based-on-triangles}.
Loop subdivision produces a piecewise linear approximation of smooth surfaces based on B-spline and multivariate spline theory.
For each edge and vertex, vertices are added in each subdivision iteration. 
The position of new vertices is computed from a convex combination of the adjacent primitives.
Considering vertex reuse, allows the merging of these memory accesses.
Figure \ref{fig:subdiv} shows results for subdividing a simplified version of the original buddha statue in this way. 
We ran one iteration of the Loop subdivision with different vertex reuse strategies on the \id{bunny}, \id{sphere} and \id{happy buddha} models.
 
We evaluated a wide variety of different parameters for batch and thread block size, and chose those producing the best results for our final consideration.
For \id{na\"ive} and \id{warp}, a batch size of 96 was used, with a block size of precisely one and two warps, respectively. 
For both \id{hash} and \id{p.hash}, we chose batches containing up to 192 indices and 64 threads per block.
Note that, for the dynamic methods, the thread block size also equals the maximum allowed number of unique indices allowed in a batch.
For \id{sort}, best results were achieved at a batch size of 768, with a block size of 256 threads. 
The Loop subdivision algorithm is arguably quite simple, and hence the cost of re-shading vertices comparably inexpensive.
However, the reduction in runtime with our vertex reuse techniques can still be as high as 26\%. 
Without exception, all vertex reuse techniques outperform the na\"{i}ve approach for the tested scenes (see Table \ref{tab:subdiv}).

\def\rtbs{\hspace{7.1pt}}
\begin{table}
	\caption{Runtimes for parallel Loop subdivision, executing on three of our input models, with different reuse methods, given in ms. 
	Our techniques achieve up to \%22 speed-up for \id{happy buddha} over naive streaming.}
	\centering
	\begin{tabular}{@{\rtbs}l@{\rtbs}@{\rtbs}r@{\rtbs}@{\rtbs}r@{\rtbs}@{\rtbs}r@{\rtbs}@{\rtbs}r@{\rtbs}@{\rtbs}r@{\rtbs}}
	\toprule
	& na\"ive & warp & hash & p.hash & sort \\
	\midrule
	bunny 			& 0.70 	& 0.58 & 0.58 & 0.59 & \textbf{0.57} \\
	sphere 			& 0.79 	& 0.68 & 0.68 & 0.69 & \textbf{0.64} \\
	happy buddha 	& 11.10 & 8.50 & \textbf{8.15} & 8.17 & 8.61\\
	\bottomrule
	\end{tabular}
	\label{tab:subdiv}
\end{table}

\def\rtbs{\hspace{5.1pt}}
\begin{table}
	\caption{Runtimes for parallel envelope creation, given in ms. Due to the particularly high shader cost, models with good vertex reuse (\id{sphere}) can achieve 3$\times$ the performance obtained with na\"{i}ve streaming alternatives.}
	\centering
	\begin{tabular}{@{\rtbs}l@{\rtbs}@{\rtbs}r@{\rtbs}@{\rtbs}r@{\rtbs}@{\rtbs}r@{\rtbs}@{\rtbs}r@{\rtbs}@{\rtbs}r@{\rtbs}}
	\toprule
	& na\"ive & warp & hash & p.hash & sort \\
	\midrule
	bunny 			& 71.73 & 29.41 & 22.21 & 21.74 & \textbf{21.85} \\
	sphere 			& 15.08 & 5.55 & \textbf{4.02} & 4.03 & 4.03 \\
	buddha 	& 5021.80 & 2112.45 & 1595.31 & 1516.15 & \textbf{1509.76} \\
	\bottomrule
	\end{tabular}
	\label{tab:envelopes}
\end{table}

\subsection{Simplification Envelopes}
The inner/outer envelopes of a mesh are defined to occupy a strict spatial sub-/superset of the input.
Resulting meshes can be used, \eg, as input to a variety of simplification algorithms, manipulation of subdivision or for conservative intersection/collision testing with tolerance
\cite{Cohen:1996:SE:237170.237220, Zhou:2007:DMS:1276377.1276491}.
An envelope is obtained by moving vertices along their vertex normal towards the \emph{inside} or the \emph{outside} of the model.
Provided that the original mesh does not contain self-intersections, an inner or outer envelope must also retain this property.
Hence, before moving each vertex, we need to determine a safe distance $\epsilon$ to ensure that no intersections will occur as a result of its transformation.
We have implemented the analytical approach presented by \citeauthor{Cohen:1996:SE:237170.237220} and ported it for parallel execution in CUDA (\citeyear{Cohen:1996:SE:237170.237220}).
Potential intersections are identified and resolved efficiently by providing an octree representation of the scene as auxiliary input.
Inner and outer envelopes for the bunny model are shown in Figure \ref{fig:envelopes}.

We configured our routine to generate outer envelopes with a target $\epsilon$ equal to 2\% of the mesh's bounding box diagonal.
We again report results with the best configuration found for each technique to yield consistently good results. 
As with subdivision, batch/block sizes for \id{na\"ive} and \id{warp} were chosen as 96/32 and 96/64, respectively.
For all dynamic methods, we found that a block size of 128 works best.
For \id{hash}, we picked a batch size of 576 indices, and used 768 indices for both \id{p.hash} and \id{sort}.
The envelope creation routine is comparably complex, and the incurred cost for each "shaded" vertex in the creation of envelopes is high:
computing an intersection-free offset for a vertex to move by requires traversing a spatial data structure, which has to be stored in global memory. 
Similarly to our experiments for rendering with high shader loads, a speed-up of more than 3$\times$ can be achieved over na\"{i}ve streaming.
Table \ref{tab:envelopes} lists reported runtimes in milliseconds for processing 2-manifold models. 

\subsection{Parallel Random Walk}
A random walk~\cite{pearson1905problem} describes a stochastic or random process, where a path is chosen on top of a graph structure or given domain, based on successive randomized steps.
Random walks are used, \eg, to simulate the paths of molecules traveling through liquids, the random search path of animals, or messages traversing through a social network.

To evaluate whether on-the-fly reuse computations can increase the performance of such random processes, we implemented a parallel walk on a discrete domain that follows a Levy flight~\cite{kleinberg2000navigation}.
We use a grid size of $256\times256$ and place \num{300000} agents on this grid.
To simulate their activity, we overlay multiple Gaussian functions on this domain.
The likelihood for agents to move a certain distance is then computed based on the activity input to the Fokker–Planck equation.
To evaluate the movements, we run through all potential moves with a maximum distance of \num{16} and keep only those \num{8} with the highest likelihood.
Then, every agent draws a random number to choose one of the stored options, whereas each is chosen with a probability proportional to their relative likelihood.

Reuse can be implemented in this scenario as follows. We encode the current agent location as a combined integer, using half of the bits for each dimension, yielding a single 32-bit word.
This number serves as a virtual \enquote{index} for the reuse computations, combining agents that are currently placed on the same grid location.
Given that the movement probability only depends on the current position, all agents with combined \enquote{indices} will see identical movement likelihoods, which can be computed only once.
The final step, which involves drawing a random number and choosing the most likely move, has to be carried out separately.

Initializing all \num{300000} agents randomly and running \num{10} simulation steps on the $256\times256$ showed that our reuse strategies can significantly increase the performance of the parallel random walk.
na\"ive, warp, hash, p.hash and sort, respectively, took \SIlist{0.30;0.10;0.09;0.13;0.10}{\milli\second} for one time step in their best configurations.
The batch sizes that achieved the best performance were rather large (1536 for dynamic batching and 576 for static batching).
At first glance, the great performance of reuse is not surprising, as the likelihood computations are rather time complex, and a high benefit can be expected for expensive vertex shaders.
However, note that the agents are also likely to significantly diverge throughout the random walk.
A further analysis revealed that a small amount of reuse already entails a significant performance gain, as the large batch sizes can still reduce the computations.


\section{Conclusion}
While the traditional solution to vertex reuse is represented by the post-transform vertex cache, caching seems to be less applicable for modern, massively parallel devices. Our simulations have shown that, even under ideal conditions, cache miss rates in a distributed environment commonly exceed 90\%. 
We have presented four inherently parallel, batch-based approaches, providing a suitable alternative to conventional caching. 
Our methods are straightforward to realize in software or hardware, and operate directly on the input buffers for indexed triangle meshes, with little to no preprocessing required.
Especially for complex shading routines, we showed that batching can achieve high reuse and increase performance by up to $3\times$ over non-reuse approaches.
We have evaluated both static and dynamic batching methods on a variety of applications and test cases. 
Due to its use of fast, warp-level communication, our static warp-voting technique is well-suited for basic shading tasks, while a dynamic, hashing-based batching approach usually performs best at high shader complexity. 
Considering that vertex shaders often exhibit low-to-medium complexity and absence of a required preprocessing step, warp-voting appears to be the recommended choice for streaming pipelines written in a compute language.

Our results are obtained from simulations and applications in CUDA, but similar approaches are straight-forward to be implemented in hardware, where communication primitives can be set up even more efficiently. 
As vertex reuse needs to interface with vertex shading, batch sizes and efficiency considerations for warp-based execution are certainly transferable into a hardware design. Furthermore, data reuse considerations are not only applicable to rendering, but also to general mesh processing and parallel graph traversal problems, where node dependencies require a similar treatment.
We hope that our implementation will help other researchers kick off the conception of novel software rendering pipelines.
Our source code is publicly available at \gitorigin.

\begin{acks}
	This research was supported by the Max Planck Center for Visual Computing and Communication, by the \grantsponsor{SP3857}{German Research Foundation (DFG)}{} grant \grantnum{SP3857}{STE 2565/1-1} and the \grantsponsor{SP357}{Austrian Science Fund (FWF)}{} grant \grantnum{SP357}{I 3007}.
	
	The Witcher game \textcopyright\ CD PROJEKT S.A.
\end{acks}

\bibliographystyle{ACM-Reference-Format}
\bibliography{vertexreuse}

\end{document}